\documentclass[review]{elsarticle}

\usepackage{lineno,hyperref}
\usepackage{mathtools}
\usepackage{amsmath}
\usepackage{bm}
\usepackage{mathrsfs}
\usepackage{amssymb}
\usepackage{euscript}
\usepackage{algorithm}
\usepackage{algorithmic}
\usepackage{graphicx}
\usepackage{multirow}
\usepackage{threeparttable}
\usepackage{subfigure}
\usepackage[Symbol]{upgreek}

\modulolinenumbers[1]

\journal{Signal Processing}

\begin{document}

\begin{frontmatter}

\title{Distributed Node-Specific Block-Diagonal LCMV
Beamforming in Wireless Acoustic Sensor Networks}

\author[mymainaddress,mysecondaryaddress]{Xinwei Guo}
\author[myfourthaddress]{Minmin Yuan}
\author[mymainaddress,mysecondaryaddress]{Chengshi Zheng\corref{mycorrespondingauthor}}
\cortext[mycorrespondingauthor]{Corresponding author}
\ead{cszheng@mail.ioa.ac.cn}
\author[mymainaddress,mysecondaryaddress]{Xiaodong Li}
\address[mymainaddress]{Key Laboratory of Noise and Vibration Research, Institute of Acoustics, Chinese Academy of Sciences,
100190, Beijing, China}
\address[mysecondaryaddress]{University of Chinese Academy of Sciences, 100049, Beijing, China}
\address[myfourthaddress]{Research Institute of Highway Ministry of Transport, 100088, Beijing, China}

\begin{abstract}
This paper derives the analytical solution of a novel distributed node-specific block-diagonal linearly constrained
minimum variance beamformer from the centralized linearly constrained minimum variance (LCMV) beamformer when considering that the noise covariance matrix is block-diagonal.
To further reduce the computational complexity of the proposed beamformer, the Sherman-Morrison-Woodbury formula is introduced to compute the inversion of noise sample covariance matrix.
By doing so, the exchanged signals can be computed with lower dimensions between nodes, where the optimal LCMV beamformer is still available at each node as if each node is to transmit its all raw sensor signal observations.
The proposed beamformer is fully distributable without imposing restrictions on the underlying network topology or scaling computational complexity, i.e., there
is no increase in the per-node complexity when new nodes are added to the networks.
Compared with state-of-the-art distributed node-specific algorithms that are often time-recursive, the proposed
beamformer exactly solves the LCMV beamformer optimally frame by frame, which has much lower computational complexity and is more robust to acoustic transfer function estimation
error and voice activity detector error. Numerous experimental results are presented
to validate the effectiveness of the proposed beamformer.
\end{abstract}


\end{frontmatter}


\section{Introduction}

Wireless acoustic sensor networks (WASNs) generally consist of several nodes, where each node has one or many sensors, a processing unit, and a wireless communication module allowing them to exchange data. Compared with the traditional and single sensor array \cite{Array},
WASNs can physically cover a wider area, which have more opportunity to select a subset of nodes close to some target sources,
and thus higher signal-to-noise ratio (SNR) and direct-to-reverberant ratio (DRR) can be expected \cite{Application,Overview}. As the next-generation technology for audio acquisition and processing, WASNs have many potential applications, such as binaural hearing aids \cite{MWF,DistributedMVDR,Hearing}, (hands-free) speech communication systems \cite{DigitalOffice,DigitalOffice2,DigitalOffice3}, and acoustic monitoring systems \cite{SP4,location1,Clustering,location2}.

In principle, all the sensor signal observations from different nodes can be transmitted to a fusion center, and then an optimal beamformer can be computed, where this approach is known as the centralized estimation \cite{LCDANSE,SP3,SP5}. The centralized estimation requires a large communication bandwidth, a large transmission power consumption at the individual nodes, and a nonnegligible computational complexity at the fusion center. However, both the power and the communication bandwidth resources in WASNs are often limited.
Furthermore, in many WASNs applications, the fusion center may be undesirable due to privacy considerations \cite{DDS1}.
A trivial solution is obtained by only utilizing the local sensor signal observations at a single node without any communication link with other nodes. Whereas, this solution cannot utilize the entire information from the WASNs and hence is only sub-optimal. A promising solution is to develop a suitable distributed approach, which often has three stages \cite{DGSC}.
At the first stage, each node processes its own sensor signal observations to obtain some compressed signals. At the second stage, only these compressed signals are
transmitted to reduce the communication bandwidth. At the last stage, a target signal is obtained by merging all these compressed signals properly.

Distributed speech enhancement algorithms can be roughly divided into two main categories: node-specific and non-node-specific. For the node-specific estimation algorithms, each node in the WASNs can estimate a different target signal, that is to say, a target source for one node may be an interfering source for another node, and vice versa. The node-specific estimation problem is intrinsic in a blind beamforming framework where the acoustic transfer functions (ATFs) between the target sound sources and the sensors are generally unknown. For these blind beamformers, some subspace estimation algorithms can be used to estimate the subspace of the ATFs \cite{Subspace1,Subspace2}, and then the target signal can be estimated as it is observed at a reference sensor. If each node in the WASNs chooses its own local sensor as reference, the spatial information of the target source can be preserved. Therefore, the node-specific estimation algorithms are preferred for many practical applications \cite{Spatial}.

Several non-node-specific speech enhancement algorithms have been presented in \cite{DDS1,DGSC}, and \cite{DDS2}, where different nodes shared a common reference sensor.
In \cite{DDS1}, each node was assumed to have one sensor and a distributed delay and sum (DDS) beamformer in a randomly connected network was proposed.
The DDS with the randomized gossip algorithm \cite{Gossip} is an iterative algorithm for solving averaging consensus problems in a distributed way, where all nodes' outputs are expected to converge to the same optimal average value. The DDS typically needs many iterations to converge to the optimal solution, as well as multiple (re)-broadcasts of the intermediary variables. The DDS is more suitable for estimating the fixed or slowly varying parameters \cite{TIDANSE}.
In \cite{DGSC}, a time-recursive distributed generalized sidelobe canceler (DGSC) was proposed for a fully connected network.
The DGSC has two components including constraints subspace and its corresponding null-space, and updates the filter coefficients during speech-absent segments. The DGSC needs to transmit
$S$-dimensional raw signal observations, with $S$ the number of target sources, in addition to the compressed signals to construct the constraints subspace component. The DGSC requires a larger communication bandwidth than those distributed algorithms transmitting only the compressed signals when one aims to get the estimation of the $S$ target source signals separately.
In \cite{DDS2}, the proposed block-diagonal LCMV (BD-LCMV) beamformer
 utilizes a set of linearly equality constraints to reduce the full-element noise sample covariance matrix to a block-diagonal form, and the imposed block-diagonal structure of the estimated sample covariance matrix results in a naturally separable objective function. Then the distributed optimal problem can be solved by
the primal-dual method of multipliers (PDMM) \cite{PDMM}. The BD-LCMV requires lots of iterations to achieve high performance, and therefore we need to make a trade-off between per-frame optimality and communication overhead in practice.

Several node-specific estimation algorithms have been proposed in \cite{MWF,DistributedMVDR,LCDANSE,TIDANSE,DANSE1,DANSE2}, and \cite{Yuan2015}, where two main criteria including the minimum mean square error (MMSE) and the minimum variance distortionless response (MVDR) are used. The mean square error (MSE) between the output signal and the desired signal comprises two components, namely the desired signal distortion and the residual noise \cite{DGSC}. The MVDR, first proposed by Capon \cite{MVDR}, minimizes the noise power at the output signal while maintaining a distortionless response towards the desired direction. Er and Cantoni \cite{LCMV} generalized the single distortionless response to a set of linear constraints, and denoted this beamformer as LCMV.

In \cite{MWF}, a distributed node-specific speech enhancement algorithm was proposed using the MMSE criterion in a 2-node network for binaural hearing aids applications. The node-specific estimation is required to preserve the auditory cues at the two ears. This method relies on the speech-distortion-weighted multichannel Wiener filter (SDW-MWF), and was referred to as the distributed MWF (DB-MWF). In \cite{DistributedMVDR}, an iterative distributed MVDR (DB-MVDR) beamformer was introduced for a similar binaural hearing aids setting. Both methods assume a single target source to obtain convergence and optimality, and are equivalent when the trade-off factor between noise reduction and distortion is zero in the SDW-MWF. A more general case was presented in \cite{TIDANSE}, \cite{DANSE1}, and \cite{DANSE2}, where multiple target sources
and $J (J\geq 2)$ nodes are considered in a so-called distributed adaptive node-specific signal estimation (DANSE) scheme. The scheme considers each node in the WASNs as a data sink, gathering the compressed signals
from other nodes, and then estimates the optimal filter coefficients in an iterative fashion. In \cite{DANSE1} and \cite{DANSE2}, the algorithms were proposed for a fully connected network and a network with a tree topology (T-DANSE), respectively. In \cite {TIDANSE}, the algorithm is topology-independent (TI-DANSE). The TI-DANSE algorithm has a slower convergence rate compared to \cite{DANSE1} and \cite{DANSE2}, and requires a larger number of frames to obtain near optimal performance \cite{DDS2}. In \cite{LCDANSE}, a distributed LCMV beamformer referred to as LC-DANSE was proposed by combining the DANSE scheme with the LCMV beamformer.
For the DANSE algorithms in \cite{TIDANSE}, \cite{DANSE1}, and \cite{DANSE2}, they attempt to align the signal components from the same source in different microphone signals. However, the alignment of the signal components is only possible when the filter length is at least twice the maximum time difference of arrival (TDOA) between all the sensors. This means that in general, the noise reduction performance degrades with increasing TDOA and a fixed filter length \cite{Hearing}. For the LC-DANSE algorithm, the raw signal observations at one node and the compressed signals from other nodes are concatenated into a new vector. In fact, the new vector still has very high dimensions, especially when the number of target sources $S$ is large, and requires a high complexity to compute the inverse of its covariance matrix. Besides, both the DANSE and the LC-DANSE algorithms are time-recursive and require multiple frames to reach optimality, which incurs a slow tracking performance \cite{DDS2}.
In \cite{Yuan2015}, each node was assumed to have more than one sensor. The recursive estimation of the inverse noise or noisy sample covariance matrix is structured as a consensus problem and is realized in a distributed manner via the randomized gossip algorithm for arbitrary topologies, similar to \cite{DDS1}. In each iteration, each node needs to transmit $M$-dimensional, with $M$ the total number of sensors in the WASNs, signals to obtain the product of the $M\times M$ inverse sample covariance matrix and the $M$-dimensional sensor signal observations. The communication cost may be higher than that of the centralized algorithm, and the convergence error accumulates across time when the aforementioned product is not accurately estimated.

In this paper, we propose a distributed node-specific block-diagonal linearly constrained minimum variance (DNBD-LCMV) beamformer. The DNBD-LCMV utilizes a set of linear equality constraints to reduce the full-element noise sample covariance matrix to a block-diagonal form and then its analytical solution can be derived from the centralized LCMV beamformer directly.
The inverse noise sample covariance matrix at each node is proposed to update by the Sherman-Morrison-Woodbury formula \cite{Sherman} and is used to compute the exchanged signals.
The proposed DNBD-LCMV can significantly reduce the number of signals exchanged between nodes, yet obtains the optimal LCMV beamformer at each node as if each node can transmit its all raw sensor signal observations. The DNBD-LCMV is fully distributable for any network topology and is completely scalable, i.e., there is no increase in the per-node computational complexity when new nodes are added to the networks. Compared with the state-of-the-art distributed node-specific algorithms,
the DNBD-LCMV exactly solves the LCMV beamformer optimally in each frame, which has much lower computational complexity and is more robust to ATF
estimation error and voice activity detector (VAD) error.

The remainder of this paper is organized as follows. In Section~\ref{Section2}, the signal model is introduced, and the centralized LCMV is also presented in this section. In Section~\ref{Section4}, the DNBD-LCMV and extension of DNBD-LCMV are shown, and its computational complexity and communication bandwidth are analyzed in Section~\ref{section5}. The experimental results are presented in Section~\ref{section6} and some conclusions are given in Section~\ref{section7}.

\section{Preliminaries\label{Section2}}
\subsection{Signal Model\label{Section21}}
We consider the WASNs with $J$ sensor nodes, where the set of nodes is denoted as $\mathcal{J}=\left\{1,\cdots,j,\cdots,J\right\}$. Each node $j$ is equipped with $M_j$ microphones and thus the total number of microphones is $M=\sum_{j=1}^JM_j$. The distributed speech enhancement problem is often formulated in the short-time Fourier transform (STFT) domain and the vector $\mathbf{y}\left(f,l\right)$ is given by
\begin{equation}
\mathbf{y}\left(f,l\right)={\left[\mathbf{y}^T_1\left(f,l\right),\cdots,\mathbf{y}^T_j\left(f,l\right),\cdots,\mathbf{y}^T_J\left(f,l\right)\right]}^T,\label{SignalMode}
\end{equation}
where $f$ denotes the frequency index, and $l$ denotes the time-frame index. $\mathbf{y}_j\left(f,l\right)$ is an $M_j\times 1$ vector consisting of locally received microphone signals at the $j$th node, and the superscript $T$ denotes the transpose operator. $\mathbf{y}\left(f,l\right)$ can be modeled as
\begin{equation}
\mathbf{y}\left(f,l\right)=\mathbf{H}\left(f,l\right)\mathbf{s}\left(f,l\right)+\mathbf{n}\left(f,l\right),
\end{equation}
where $\mathbf{s}\left(f,l\right)$ is a signal vector containing $S$ speech sources, $\mathbf{n}\left(f,l\right)$ is a noise vector, and
\begin{equation}
\mathbf{H}\left(f,l\right)={\left[ \mathbf{H}^T_1\left(f,l\right),\cdots,\mathbf{H}^T_j\left(f,l\right),\cdots,\mathbf{H}^T_J\left(f,l\right)\right]}^T
\end{equation}
is a full-rank $M\times S$ ATF matrix. In particularly, $\mathbf{H}_j\left(f,l\right)$ is the ATF matrix between the $S$ speech sources and the $M_j$ microphones at the $j$th node. In the following, $\mathbf{H}\left(f,l\right)$ can be approximated as time-invariant in each frame and all derivations refer to a single frequency bin.
The frame index $l$ and the frequency index $f$ will be omitted when no confusion arises.

\subsection{Centralized LCMV Beamforming\label{Section22}}

For the centralized LCMV beamformer, node $j$ applies a $M$-dimensional estimator $\mathbf{w}_j$ to the
$M$-dimensional microphone signals $\mathbf{y}$ to obtain the node-specific output $d_j=\mathbf{w}^H_j\mathbf{y}$, where the superscript $H$ denotes the conjugate transpose operator. $\mathbf{w}_j$ can be obtained
from the following general optimization problem \cite{BF}\cite{Frost}
\begin{equation}
\begin{gathered}
\min \limits_{\mathbf{w}_j}~~\mathbf{w}^H_j\mathbf{R}_{nn}\mathbf{w}_j,\\
\mathrm{s.t.}~~\mathbf{w}^H_j\mathbf{H}=\mathbf{g}^H_j, \label{Problem}
\end{gathered}
\end{equation}
where $\mathbf{R}_{nn}=E\left\{\mathbf{n}\mathbf{n}^H\right\}$ is the noise covariance matrix and $E\{\cdot\}$ denotes the expected value operator.
$\mathbf{g}_j$ is an $S\times 1$ desired response vector for the $S$ speech sources. Its entries usually consist of ones and zeros to preserve the target sources and eliminate other interfering sources simultaneously. The solution of (\ref{Problem}) can be given by
\begin{equation}
\mathbf{w}_j=\mathbf{R}^{-1}_{nn}\mathbf{H}\left(\mathbf{H}^H\mathbf{R}^{-1}_{nn}\mathbf{H}\right)^{-1}\mathbf{g}_j.\label{CenSolution1}
\end{equation}
The node-specific output can be expressed as
\begin{equation}
\begin{split}
d_j&=\mathbf{w}^H_j\mathbf{y}\\
   &=\mathbf{g}^H_j\left(\mathbf{H}^H\mathbf{R}^{-1}_{nn}\mathbf{H}\right)^{-1}\mathbf{H}^H\mathbf{R}^{-1}_{nn}\left(\mathbf{H}\mathbf{s}+\mathbf{n}\right)\\
   &=\sum_{k=1}^Sg^{*}_j\left(k\right)s\left(k\right)+\mathbf{w}^H_j\mathbf{n},\label{CenSolution2}
\end{split}
\end{equation}
where $g_j\left(k\right)$ and $s\left(k\right)$ are the $k$th entries of $\mathbf{g}_j$ and $\mathbf{s}$, respectively. The superscript ``$*$'' marker denotes the conjugate operator.

Equations (\ref{CenSolution1}) and (\ref{CenSolution2}) require each node to have access to all microphone signals $\mathbf{y}$ to estimate $\mathbf{R}_{nn}$ and then obtain $d_j$. Therefore, all the locally received signals $\mathbf{y}_j$ at the $j$th node need to be
transmitted, which results in a large communication bandwidth and a large transmission power in the WASNs. Besides, the computational power grows dramatically with the increase of $M$ when computing the inversion of $\mathbf{R}_{nn}$.

\section{Method\label{Section4}}
In the previous section, it is assumed that each node transmits all its microphone signals to every other node in the WASNs such that each node can compute (\ref{CenSolution1}) and (\ref{CenSolution2}).
We now look, instead, to the case where each node only transmits a linearly compressed version of its microphone signals by means of the distributed node-specific block-diagonal LCMV (DNBD-LCMV) beamformer.
For the sake of an easy exposition, we first assume that the ATF matrix $\mathbf{H}$ is known, and describe the
DNBD-LCMV for a fully connected network, where each node is able to directly communicate with every other node in the WASNs. Then, all derivations are extended to a blind beamforming framework and any topology, similar to \cite{DDS2}.

\subsection{DNBD-LCMV with Known ATF Matrix}
If the reverberation time of a room is moderate or large enough and/or the noise is far away from the microphones, i.e., the distance between the noise and the microphones is larger than the reverberation radius, its reverberant sound field is diffuse, homogenous, and isotropic \cite{DDS2} and \cite{Sinc}. In this case,
the normalized correlation ${\Omega}_{m,m_1}\left(f\right)$ between two microphones $m$ and $m_1$ with distance ${\lambda}_{m,m_1}$ at a frequency $f$ can be given by
\begin{equation}
{\Omega}_{m,m_1}\left(f\right)=\frac{\mathrm{sin}\left(2\pi f{\lambda}_{m,m_1}/c\right)}{2\pi f{\lambda}_{m,m_1}/c},\label{sinc}
\end{equation}
where $c=343~\mathrm{m/s}$ is the sound speed.
The correlation can be roughly divided into two frequency regions: one highly correlated
at low frequencies and the other much less correlated at high frequencies. The boundary between the two regions occurs at the first zero-crossing frequency $f_c=c/\left(2{\lambda}_{m,m_1}\right)$. When the distance ${\lambda}_{m,m_1}$ is large, the frequency
$f_c$ is small. For example, $f_c$ equals 171.5 Hz for ${\lambda}_{m,m_1}=1$ m.

For the WASNs, the microphones within a node are often nearby, whereas the microphones from different nodes are further away. The noise can be assumed to be uncorrelated across the different nodes \cite{DDS2}, and then $\mathbf{R}_{nn}\left(l\right)$ has the following block-diagonal form approximately
\begin{equation}
\begin{split}
\mathbf{R}_{nn}\left(l\right)&=\mathrm{Blockdiag}\left(\mathbf{\Delta}_{nn,1}\left(l\right),\cdots,\mathbf{\Delta}_{nn,j}\left(l\right),\cdots,
                                \mathbf{\Delta}_{nn,J}\left(l\right)\right)  \\
                             &=\begin{bmatrix}
\mathbf{\Delta}_{nn,1}\left(l\right) & \cdots & \mathbf{0}_{M_1\times M_j}& \cdots & \mathbf{0}_{M_1\times M_J}\\
          \vdots   & \ddots & \vdots & \vdots & \vdots\\
         \mathbf{0}_{M_j\times M_1} & \cdots &\mathbf{\Delta}_{nn,j}\left(l\right) & \cdots& \mathbf{0}_{M_j\times M_J}   \\
         \vdots &  \vdots&   \vdots   &  \ddots & \vdots\\
         \mathbf{0}_{M_J\times M_1}& \cdots & \mathbf{0}_{M_J\times M_j} & \cdots & \mathbf{\Delta}_{nn,J}\left(l\right)
\end{bmatrix},\label{blockmatrix}
\end{split}
\end{equation}
where $\mathbf{\Delta}_{nn,j}\left(l\right)$ is the noise covariance matrix at the $j$th node, and $\mathbf{0}_{M_{j_1}\times M_{j_2}}$ is an $M_{j_1}\times M_{j_2}$ null matrix.

Note that the noise covariance matrix used in the existing node-specific algorithms, such as \cite{LCDANSE,TIDANSE,DANSE1,DANSE2}, and \cite{Yuan2015}, is full-element. In the non-node-specific algorithm BD-LCMV \cite{DDS2}, the block-diagonal $\mathbf{R}_{nn}\left(l\right)$ is adopted and the weight vector associated with the $j$th node is given by
\begin{equation}
\mathbf{w}_j(l)=\mathbf{\Delta}^{-1}_{nn,j}\left(l\right)\mathbf{H}_j\bm{\mu}\left(l\right),
\end{equation}
where $\bm{\mu}\left(l\right)$ is a Lagrange multiplier shared by all nodes. Then the dual optimization problem is introduced to compute the optimal $\bm{\mu}\left(l\right)$ and is solved by PDMM. Finally, the signal $e_j(l)=\mathbf{w}^H_j(l)\mathbf{y}_j(l)$ is exchanged between nodes and the output $d_j(l)$ is obtained by summing all $e_j(l)$.

In this paper, the proposed beamformer has a completely new scheme. From (\ref{CenSolution1}) and (\ref{blockmatrix}), the weight vector corresponding to the $l$th-frame microphone signals can be expressed by
\begin{equation}
\begin{split}
\mathbf{w}_j(l)&=\mathbf{R}^{-1}_{nn}\left(l\right)\mathbf{H}\left(\mathbf{H}^H\mathbf{R}^{-1}_{nn}\left(l\right)\mathbf{H}\right)^{-1}\mathbf{g}_j,\\
            &=\begin{bmatrix}
\bm{\Delta}^{-1}_{nn,1}\left(l\right)\mathbf{H}_1\\
\vdots\\
\bm{\Delta}^{-1}_{nn,j}\left(l\right)\mathbf{H}_j\\
\vdots\\
\bm{\Delta}^{-1}_{nn,J}\left(l\right)\mathbf{H}_J
\end{bmatrix}\left(\sum_{j_1=1}^J\mathbf{H}^H_{j_1}\bm{\Delta}^{-1}_{nn,j_1}\left(l\right)\mathbf{H}_{j_1}\right)^{-1}\mathbf{g}_j.\\ \label{BDWeight}
\end{split}
\end{equation}

The $S$-dimensional compressed signals $\mathbf{z}_j\left(l\right)$ and the $S\times S$ matrix $\mathbf{D}_j\left(l\right)$ related to the $j$th node are defined as
\begin{equation}
\begin{gathered}
\mathbf{z}_j\left(l\right)=\mathbf{H}^H_j\mathbf{\Delta}^{-1}_{nn,j}\left(l\right)\mathbf{y}_j\left(l\right),\\
\mathbf{D}_j\left(l\right)=\mathbf{H}^H_j\mathbf{\Delta}^{-1}_{nn,j}\left(l\right)\mathbf{H}_j. \label{CompressedSignal}
\end{gathered}
\end{equation}
From (\ref{CenSolution2}) and (\ref{BDWeight}), the node-specific output can be rewritten as
\begin{equation}
\begin{split}
d_j\left(l\right)&=\mathbf{w}^H_j\left(l\right)\mathbf{y}\left(l\right)\\
      &=\mathbf{g}^H_j\tilde{\mathbf{z}}\left(l\right),\label{Tequation}
\end{split}
\end{equation}
where $\tilde{\mathbf{z}}\left(l\right)$ is the product of $\mathbf{D}^{-1}\left(l\right)$ and $\mathbf{z}\left(l\right)$,
\begin{equation}
\tilde{\mathbf{z}}\left(l\right)=\mathbf{D}^{-1}\left(l\right)\mathbf{z}\left(l\right). \label{Multiplier1}
\end{equation}
In particularly, $\mathbf{D}\left(l\right)$ and $\mathbf{z}\left(l\right)$ are obtained by
summing all $\mathbf{z}_j\left(l\right)$ and all $\mathbf{D}_j\left(l\right)$ separately,
\begin{equation}
\mathbf{D}\left(l\right)=\sum_{j=1}^J\mathbf{D}_{j_1}\left(l\right),~\mathbf{z}\left(l\right)=\sum_{j=1}^J\mathbf{z}_{j_2}\left(l\right).\label{summation}
\end{equation}

The above derivations assume that the noise covariance matrix $\mathbf{\Delta}_{nn,j}\left(l\right)$
at the $j$th node is perfectly known. However, for practical applications, $\mathbf{\Delta}_{nn,j}\left(l\right)$ is unknown and needs to be estimated from the noisy observations. This often requires a hard or soft VAD to determine whether the speakers are present or not. It is noted that the design of a VAD mechanism is a hot research topic on its own, and is out of the scope of this paper \cite{DGSC,TIDANSE,SP6}. In the following, two methods including non-recursive (moving average) smoothing and first-order recursive smoothing are considered to estimate $\mathbf{\Delta}_{nn,j}\left(l\right)$.

\subsubsection{Non-Recursive Smoothing Method\label{Section411}}

For the non-recursive smoothing method, the set of microphone signals frames and the set of noise-only frames for the current block of microphone signals are denoted by
$\mathcal{L}_y$ and $\mathcal{L}_n \left(\mathcal{L}_n\subseteq \mathcal{L}_y\right)$, respectively. The block has $\lvert \mathcal{L}_y\rvert$ frames microphone signals and $\lvert \cdot \rvert$ denotes the cardinality of a set. The noise covariance matrix at the $j$th node can be estimated using the set of noise-only frames,
\begin{equation}
\mathbf{\Delta}_{nn,j}\left(l\right)\approx \widehat{\mathbf{\Delta}}_{nn,j}\left(l\right)=\frac{1}{\lvert \mathcal{L}_n \rvert}\sum_{l_n\in \mathcal{L}_n}\mathbf{y}_j\left(l_n\right)\mathbf{y}^H_j\left(l_n\right). \label{NonRecursive}
\end{equation}
Rigorously, all the estimated values need to use ``$~\widehat{}~$'' to distinguish them from their
true values. Analogously to $\mathbf{\Delta}_{nn,j}\left(l\right)$ in (\ref{NonRecursive}),
we omit ``$~\widehat{}~$'' in the following for the sake of brevity when no confusion arises.
For the current block of microphone signals, $\mathbf{\Delta}_{nn,j}\left(l\right)$ only needs to be estimated once. From (\ref{CompressedSignal}), $\mathbf{D}_j\left(l\right)$ also needs to be estimated once.

\begin{figure}[t]
\centering
\includegraphics[width=90mm]{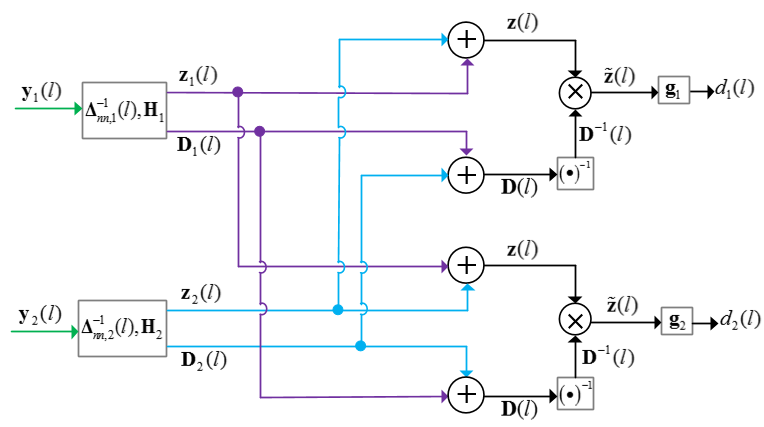}
\caption{The scheme of the DNBD-LCMV with the non-recursive smoothing method for the current
block of microphone signals ($J=2$). Each node $j$ computes its node-specific output $d_j\left(l\right)$ using its own $S$-dimensional compressed signals $\mathbf{z}_j\left(l\right)$ and $S$-dimensional compressed signals transmitted by the other node ($\mathbf{D}_j\left(l\right)$ only needs to be transmitted once for the current block of microphone signals).}\label{FlowChart}
\end{figure}

We assume that each block has $B$ frames microphone signals, i.e., $\lvert \mathcal{L}_y\rvert=B$. The scheme of the DNBD-LCMV with the
non-recursive smoothing method is shown in Fig.~\ref{FlowChart} and it
consists of the following steps:\\

1) Initialize: $i\gets 0$.

2) Each node $j\in \mathcal{J}$ performs the following operation cycle:

\hangafter=1
\setlength{\hangindent}{3em}
\setlength{\parindent}{2em}{
$\bullet$ Collect the current block of microphone signals $\mathbf{y}_j\left(l\right),l\in \mathcal{L}_y=\{iB,iB+1,\cdots,\left(i+1\right)B-1\}$.}

\hangafter=1
\setlength{\hangindent}{3em}
\setlength{\parindent}{2em}{
$\bullet$ For the current block of microphone signals,
$\mathbf{\Delta}_{nn,j}\left(l\right)$ is estimated with (\ref{NonRecursive}).}

\hangafter=1
\setlength{\hangindent}{3em}
\setlength{\parindent}{2em}{
$\bullet$ $\mathbf{H}^H_j\mathbf{\Delta}^{-1}_{nn,j}\left(l\right)$ is applied to $M_j$-dimensional microphone signals $\mathbf{y}_j\left(l\right)$ to obtain the $S$-dimensional compressed signals $\mathbf{z}_j\left(l\right)$ with (\ref{CompressedSignal}). Besides, $\mathbf{D}_j\left(l\right)$ can also be obtained.}

\hangafter=1
\setlength{\hangindent}{3em}
\setlength{\parindent}{2em}{
$\bullet$ $\mathbf{z}_j(l)$ and $\mathbf{D}_j\left(l\right)$ are transmitted. In particularly, $\mathbf{D}_j\left(l\right)$ only needs to be transmitted once for the current block of microphone signals.}

\hangafter=1
\setlength{\hangindent}{3em}
\setlength{\parindent}{2em}{
$\bullet$ Each node can have access to all $\mathbf{z}_j(l)$ and $\mathbf{D}_j\left(l\right)$, and
computes $\mathbf{z}(l)$ and $\mathbf{D}\left(l\right)$ with (\ref{summation}). Then, the inverse matrix $\mathbf{D}^{-1}\left(l\right)$ is multiplied with $\mathbf{z}(l)$ to obtain $\tilde{\mathbf{z}}(l)$.
Finally, $\mathbf{g}_j$ is applied to $\tilde{\mathbf{z}}(l)$ to obtain $d_j(l)$.}

\setlength{\parindent}{1em}
3) $i\gets i+1$.

4) Return to step 2).\\

From Fig.~\ref{FlowChart}, for each block of microphone signals, $\mathbf{D}_j\left(l\right)$ only needs to be transmitted once. Since $\mathbf{D}_j\left(l\right)$ is a Hermitian matrix and its main diagonal entries are real numbers, the transmission of $\mathbf{D}_j\left(l\right)$ results in a total of $S^2$ transmitted real numbers for each frequency bin per block. The transmission cost is slight and can be neglected.

\subsubsection{First-Order Recursive Smoothing Method}

For the first-order recursive smoothing method, the noise sample covariance matrix $\mathbf{\Delta}_{nn,j}\left(l\right)$ at the $j$th node is updated by \cite{MWF1}
\begin{equation}
\begin{split}
\mathbf{\Delta}_{nn,j}\left(l\right)&=\left(1-\mathcal{P}_j\left(l\right)\right)\big(\alpha \mathbf{\Delta}_{nn,j}\left(l-1\right)+\left(1-\alpha\right)\mathbf{y}_j\left(l\right)\mathbf{y}^H_j\left(l\right)\big)
\\&\quad+\mathcal{P}_j\left(l\right) \mathbf{\Delta}_{nn,j}\left(l-1\right),
\end{split}
\end{equation}
where $\mathcal{P}_j\left(l\right)=1$ when speech component is detected in
the $l$th-frame microphone signals $\mathbf{y}_j\left(l\right)$; and $\mathcal{P}_j\left(l\right)=0$, otherwise. When the binary VAD decision is replaced by a soft speech presence probability (SPP) \cite{SPP}, $\mathcal{P}_j\left(l\right)$ can vary from 0 to 1, which will not be further considered here.
$\alpha$ is a forgetting factor ranging from 0 to 1.

When $\mathcal{P}_j\left(l\right)=1$, $\mathbf{\Delta}_{nn,j}\left(l\right)$ is not updated, i.e., $\mathbf{\Delta}_{nn,j}\left(l\right)=\mathbf{\Delta}_{nn,j}\left(l-1\right)$.
Similar to Section~\ref{Section411}, only the transmission of $\mathbf{z}_j(l)$ is required. While, for $\mathcal{P}_j\left(l\right)=0$, i.e., $\mathbf{y}_j(l)=\mathbf{n}_j(l)$, the current noise-only frame is included to update $\mathbf{\Delta}_{nn,j}\left(l\right)$ by the following equation
\begin{equation}
\mathbf{\Delta}_{nn,j}\left(l\right)=\alpha \mathbf{\Delta}_{nn,j}\left(l-1\right)+\left(1-\alpha\right)\mathbf{y}_j\left(l\right)\mathbf{y}^H_j\left(l\right). \label{SCM}
\end{equation}
A naive solution is obtained by transmitting $\mathbf{D}_j(l)$ and $\mathbf{z}_j(l)$ for each noise-only frame, where a total of $(S^2+2S)$ transmitted real numbers and the inversion operation of $\mathbf{\Delta}_{nn,j}\left(l\right)$ are required. The communication cost and computational load are huge and need to be further reduced.

With the help of the Sherman-Morrison-Woodbury formula \cite{Sherman}, the inversion of
$\mathbf{\Delta}_{nn,j}\left(l\right)$ in (\ref{SCM}) can be expressed by
\begin{equation}
\begin{gathered}
\mathbf{\Delta}^{-1}_{nn,j}\left(l\right)=\frac{1}{\alpha}\Bigg(\mathbf{\Delta}^{-1}_{nn,j}\left(l-1\right)
-\frac{\mathbf{\Delta}^{-1}_{nn,j}\left(l-1\right)\mathbf{y}_j\left(l\right)\mathbf{y}^H_j\left(l\right)\mathbf{\Delta}^{-1}_{nn,j}\left(l-1\right)}{\alpha/\left(1-\alpha\right)+\mathbf{y}^H_j\left(l\right)\mathbf{\Delta}^{-1}_{nn,j}\left(l-1\right)\mathbf{y}_j\left(l\right)}\Bigg).
\label{Formula}
\end{gathered}
\end{equation}
We further define $\bar{\mathbf{c}}_j\left(l\right),c_j\left(l\right)$, and $\tilde{\mathbf{c}}_j\left(l\right)$ as follows:
\begin{equation}
\begin{gathered}
\bar{\mathbf{c}}_j\left(l\right)=\mathbf{H}^H_j\mathbf{\Delta}^{-1}_{nn,j}\left(l-1\right)\mathbf{y}_j\left(l\right),\\
c_j\left(l\right)=\mathbf{y}^H_j\left(l\right)\mathbf{\Delta}^{-1}_{nn,j}\left(l-1\right)\mathbf{y}_j\left(l\right),\\
\tilde{\mathbf{c}}_j\left(l\right)=\mathbf{\Delta}^{-1}_{nn,j}\left(l-1\right)\mathbf{y}_j\left(l\right).\label{definition}
\end{gathered}
\end{equation}

By substituting (\ref{Formula}) and (\ref{definition}) into (\ref{CompressedSignal}), we get
\begin{equation}
\begin{split}
\mathbf{D}_j\left(l\right)&=\mathbf{H}^H_j\mathbf{\Delta}^{-1}_{nn,j}\left(l\right)\mathbf{H}_j\\
               &=\frac{1}{\alpha}\left(\mathbf{D}_j\left(l-1\right)-\frac{\bar{\mathbf{c}}_j\left(l\right)\bar{\mathbf{c}}^H_j\left(l\right)}{\alpha/(1-\alpha)+c_j\left(l\right)}\right),\label{Reconstruction1}
\end{split}
\end{equation}
and
\begin{equation}
\begin{split}
\mathbf{z}_j\left(l\right)&=\mathbf{H}^H_j\mathbf{\Delta}^{-1}_{nn,j}\left(l\right)\mathbf{y}_j\left(l\right)\\
               &=\frac{\bar{\mathbf{c}}_j\left(l\right)}{\alpha}\left(1-\frac{c_j\left(l\right)}{\alpha/\left(1-\alpha\right)+c_j(l)}\right).\label{Reconstruction2}
\end{split}
\end{equation}
By substituting $\left(\ref{definition}\right)$ into $\left(\ref{Formula}\right)$
, $\left(\ref{Formula}\right)$ can be further written as
\begin{equation}
\mathbf{\Delta}^{-1}_{nn,j}\left(l\right)=\frac{1}{\alpha}\left(\mathbf{\Delta}^{-1}_{nn,j}\left(l-1\right)
-\frac{\tilde{\mathbf{c}}_j\left(l\right)\tilde{\mathbf{c}}^H_j\left(l\right)}{\alpha/\left(1-\alpha\right)+c_j\left(l\right)}\right).\label{Reconstruction3}
\end{equation}

For the noise-only frame, i.e., $\mathcal{P}_j\left(l\right)=0$, from (\ref{Reconstruction1}), (\ref{Reconstruction2}), and (\ref{Reconstruction3}), the scheme
of the DNBD-LCMV with the first-order recursive smoothing method is shown in Fig.~\ref{FlowChart2}, and it consists of the following steps:\\

1) Each node $j\in \mathcal{J}$ performs the following operation cycle:

\hangafter=1
\setlength{\hangindent}{3em}
\setlength{\parindent}{2em}{$\bullet$ The inverse matrix $\mathbf{\Delta}^{-1}_{nn,j}\left(l-1\right)$ is applied to $M_j$-dimensional microphone signals $\mathbf{y}_j\left(l\right)$ to obtain the $M_j$-dimensional vector $\tilde{\mathbf{c}}_j\left(l\right)$ and the real number $c_j\left(l\right)$ with (\ref{definition}), where the $S$-dimensional vector $\bar{\mathbf{c}}_j\left(l\right)$ can also be obtained by using $\mathbf{H}_j$. Then, $\mathbf{\Delta}^{-1}_{nn,j}\left(l\right)$ is estimated with (\ref{Reconstruction3}).}

\hangafter=1
\setlength{\hangindent}{3em}
\setlength{\parindent}{2em}{$\bullet$ $\bar{\mathbf{c}}_j\left(l\right)$ and $c_j\left(l\right)$ are transmitted.

\hangafter=1
\setlength{\hangindent}{3em}
\setlength{\parindent}{2em}{$\bullet$ Each node can have access to $\bar{\mathbf{c}}_j\left(l\right)$ and $c_j\left(l\right)$, and reconstructs $\mathbf{D}_j\left(l\right)$ and $\mathbf{z}_j\left(l\right)$ with (\ref{Reconstruction1}) and (\ref{Reconstruction2}), respectively. Then, $\mathbf{D}\left(l\right)$ and $\mathbf{z}\left(l\right)$ are computed with (\ref{summation}). The inverse matrix $\mathbf{D}^{-1}\left(l\right)$ is multiplied with $\mathbf{z}\left(l\right)$ to obtain $\tilde{\mathbf{z}}\left(l\right)$.
Finally, $\mathbf{g}_j$ is applied to $\tilde{\mathbf{z}}\left(l\right)$ to obtain $d_j\left(l\right)$.}

\setlength{\parindent}{1em}
2) $ l\gets l+1$.

3) Return to step 1).\\

From Fig.~\ref{FlowChart2}, for each noise-only frame, the $S$-dimensional vector $\bar{\mathbf{c}}_j\left(l\right)$ and the real number $c_j\left(l\right)$ are transmitted to reconstruct the $S\times S$ matrix $\mathbf{D}_j\left(l\right)$ and the $S$-dimensional vector $\mathbf{z}_j\left(l\right)$ at other nodes instead of the transmission of $\mathbf{D}_j\left(l\right)$ and $\mathbf{z}_j\left(l\right)$.
Besides, the inversion operation of $\mathbf{\Delta}_{nn,j}\left(l\right)$ is only required at the beginning. The communication cost and computational load are greatly reduced compared
to the naive solution mentioned above.

\begin{figure*}[t]
\centering
\includegraphics[width=100mm]{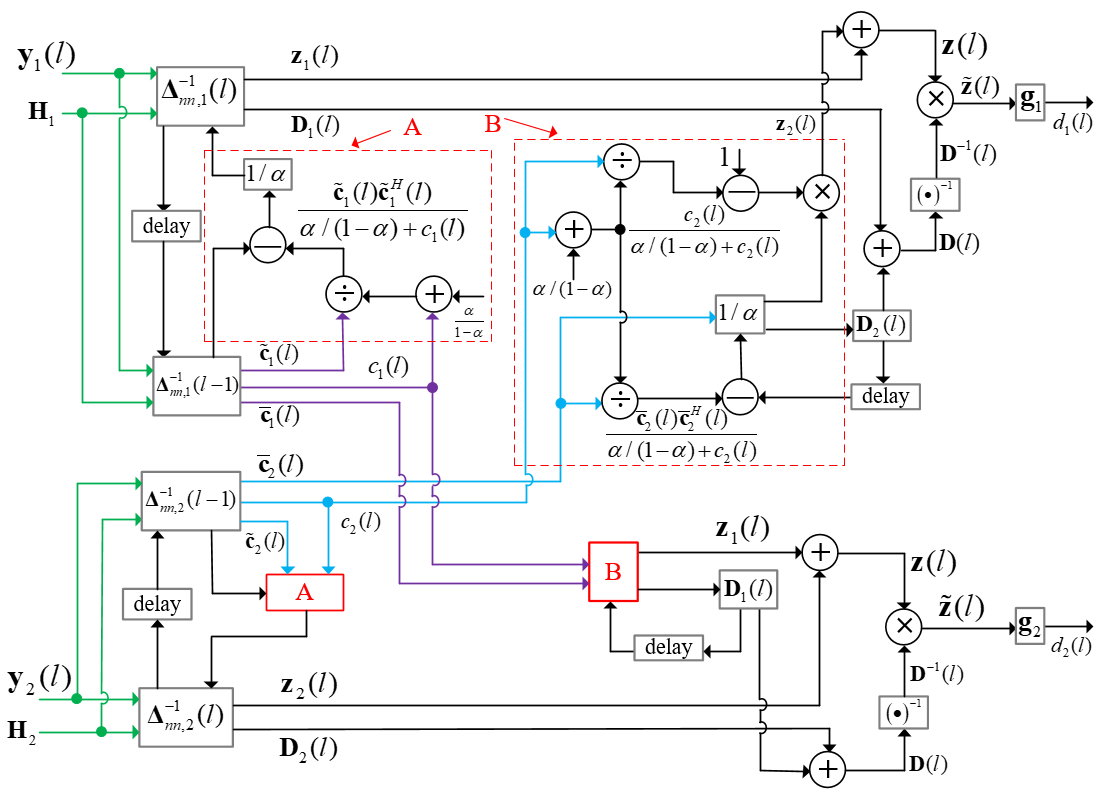}
\caption{The scheme of the DNBD-LCMV with the first-order recursive smoothing method ($J=2,\mathcal{P}_j\left(l\right)=0$). For each node $j$, the $S$-dimensional vector $\bar{\mathbf{c}}_j\left(l\right)$ and the real number $c_j\left(l\right)$ are transmitted
to reconstruct $\mathbf{D}_j\left(l\right)$ and $\mathbf{z}_j\left(l\right)$ with (\ref{Reconstruction1}) and (\ref{Reconstruction2}), respectively, at the other node.
The inverse matrix $\mathbf{\Delta}^{-1}_{nn,j}\left(l\right)$ is updated with (\ref{Reconstruction3}). Each node $j$ computes its node-specific output $d_j\left(l\right)$ use its own $S$-dimensional compressed signal
$\mathbf{z}_j\left(l\right)$ and the reconstructed signal $\mathbf{z}_q\left(l\right),q\in \mathcal{J}\setminus \{j\}$, where $\mathbf{D}_j\left(l\right)$ and the reconstructed matrix $\mathbf{D}_q\left(l\right)$ are also used.}\label{FlowChart2}
\end{figure*}

\subsection{DNBD-LCMV with Unknown ATF Matrix}
In general, the ATF matrix $\mathbf{H}$ is unknown and needs to be estimated online. Some subspace estimation algorithms can be used to estimate the column space
of $\mathbf{H}$ \cite{Subspace1,Subspace2}. Define $\mathbf{Q}=\left[\mathbf{q}_1,\cdots,\mathbf{q}_S\right]$ to be any basis that spans the column space of $\mathbf{H}$,
\begin{equation}
\mathbf{H}=\mathbf{Q}\mathbf{\Theta},
\end{equation}
where $\mathbf{\Theta}$ is an $S\times S$ matrix comprised of the projection coefficients of the original ATFs on the basis vectors.

When the speakers only change their positions slowly with respect to their initial positions, such as teleconferencing, we can estimate the column space $\mathbf{Q}$ at the initial stage (e.g., in a centralized way) \cite{DDS2}. This may cause some estimation error of $\mathbf{Q}$ if the speakers have some slight movements and therefore robust beamformer is preferred.

The goal for node $j$ is to estimate the target signal from the signal vector $\mathbf{s}$ as observed by one of node $j$'s microphones, referred to as the reference microphone. Without loss of generality, the first microphone of each node is chosen as the reference microphone. For node $j$, the index of its reference microphone is equal to $m=1+\sum_{q=1}^{\left(j-1\right)}M_{q}$.
Accordingly, the weight vector in (\ref{CenSolution1}) can be rewritten as
\begin{equation}
\begin{gathered}
\bar{\mathbf{w}}_j=\mathbf{R}^{-1}_{nn}\mathbf{Q}\left(\mathbf{Q}^H\mathbf{R}^{-1}_{nn}\mathbf{Q}\right)^{-1}\bar{\mathbf{g}}_j,\\
\bar{g}_j\left(k\right)=g_j\left(k\right)Q^{*}\left(m,k\right),\\
\label{UnknownMatrix}
\end{gathered}
\end{equation}
where $Q\left(m,k\right)$ is the entry in the $m$th row and $k$th column of $\mathbf{Q}$. $\bar{g}_j\left(k\right)$ and $g_j\left(k\right)$ are the $k$th entry of $\bar{\mathbf{g}}_j$ and that of $\mathbf{g}_j$, respectively.
The node-specific output in (\ref{CenSolution2}) is modified by the following equation \cite{LCDANSE}
\begin{equation}
\begin{split}
\bar{d}_j&=\bar{\mathbf{w}}^H_j\mathbf{y}\\
  &=\bar{\mathbf{g}}^H_j\left(\mathbf{Q}^H\mathbf{R}^{-1}_{nn}\mathbf{Q}\right)^{-1}\mathbf{Q}^H\mathbf{R}^{-1}_{nn}\left(\mathbf{Q}\mathbf{\Theta}\mathbf{s}+\mathbf{n}\right)\\
   &=\sum_{k=1}^{S}g^{*}_j\left(k\right)H\left(m,k\right)s\left(k\right)+\bar{\mathbf{w}}^H_j\mathbf{n},
\end{split}
\end{equation}
where $H\left(m,k\right)$ is the entry in the $m$th row and $k$th column of $\mathbf{H}$.

It is obvious that different nodes have different desired response vectors due to different reference microphones, i.e., $\bar{\mathbf{g}}_j\not=\bar{\mathbf{g}}_{q}$ with $j\not=q$. Therefore, each node can extract a node-specific target signal based on node-specific reference microphone to preserve the spatial information of the target source, such as time difference cues, which are very important in target source localization.

\subsection{Extension of DNBD-LCMV}

When the noise covariance matrix $\mathbf{\Delta}_{nn,j}$ in (\ref{blockmatrix}) is replaced by the noisy covariance matrix $\mathbf{\Delta}_{yy,j}$, that is to say,
\begin{equation}
\mathbf{R}_{yy}=\mathrm{Blockdiag}\left(\mathbf{\Delta}_{yy,1},\cdots,\mathbf{\Delta}_{yy,j},\cdots,
                                \mathbf{\Delta}_{yy,J}\right), \label{blockmatrix2}
\end{equation}
one can obtain the distributed node-specific block-diagonal linearly constrained minimum power (DNBD-LCMP) beamformer, which can be given by
\begin{equation}
\bar{\mathbf{w}}^{'}_j=\mathbf{R}^{-1}_{yy}\mathbf{Q}\left(\mathbf{Q}^H\mathbf{R}^{-1}_{yy}\mathbf{Q}\right)^{-1}\bar{\mathbf{g}}_j.
\end{equation}
where this beamformer does not require an estimate of the noise covariance matrix.

Particularly, when $\mathbf{\Delta}_{nn,j}$ is an identity matrix $\mathbf{I}_j$, the DNBD-LCMV becomes the distributed node-specific delay and sum (DNDS) beamformer
\begin{equation}
\bar{\mathbf{w}}^{''}_j=\mathbf{Q}\left(\mathbf{Q}^H\mathbf{Q}\right)^{-1}\bar{\mathbf{g}}_j. \label{DNDS}
\end{equation}
where this beamformer only needs to be calculated once at the beginning. However, it cannot control the sound sources that are not included in $\bar{\mathbf{g}}_j$.

Similar to \cite{DDS2}, the proposed beamformers including DNBD-LCMV/DNBD-LCMP, and DNDS can be implemented in the WASNs with arbitrary topologies. This can be achieved by a slight modification to the data transmission process of each node, where each node aggregates its transmitted data including $\mathbf{z}_j\left(l\right)$ and $\mathbf{D}_j\left(l\right)$ with the transmitted data from its neighbors. This will allow for the transmitted data to disperse through the WASNs by means of an in-network summation, and is demonstrated for the tree topology, as shown in Fig.~\ref{Tree}.
\begin{figure}[t]
\centering
\includegraphics[width=70mm]{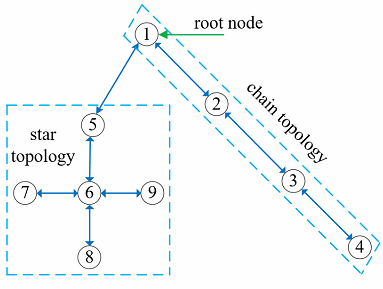}
\caption{The tree topology with two sub-topologies: chain and star.}\label{Tree}
\end{figure}

We denote $\mathcal{N}_j$ as the set of neighbors of node $j$ in the topology with node $j$ excluded. After the tree is formed with any tree formation algorithm \cite{SpanTree,SpanTree2},
one arbitrary node is assigned as the root node and the nodes only communicate with their neighbors (for example, $\mathcal{N}_6=\{5,7,8,9\}$ and node 1 is the root node in Fig.~\ref{Tree}). The following data-driven signal flow is executed for each block of microphone signals (non-recursive smoothing method is considered below):

\hangafter=1
\setlength{\hangindent}{2.2em}
1) Any leaf node, i.e., a non-root node $q$ having only a single neighbor, can immediately fire and transmits $\mathbf{z}_q\left(l\right)$ and $\mathbf{D}_q\left(l\right)$ to its single neighbor (toward the root node).
Any non-root node $j$ with more than a single neighbor waits until it has received the transmitted data from all its  neighbors except for a single neighbor that has yet to fire, say node $j^{'}$, and then computes the following sum
\begin{equation}
\begin{gathered}
\check{\mathbf{z}}_j\left(l\right)=\mathbf{z}_j\left(l\right)+\sum_{q\in \mathcal{N}_j\setminus j^{'}} \mathbf{z}_q\left(l\right),\\
\check{\mathbf{D}}_j\left(l\right)=\mathbf{D}_j\left(l\right)+\sum_{q\in \mathcal{N}_j\setminus j^{'}} \mathbf{D}_q\left(l\right).
\end{gathered}
\end{equation}
Next, $\check{\mathbf{z}}_j\left(l\right)$ and $\check{\mathbf{D}}_j\left(l\right)$ are transmitted to node $j^{'}$ (toward the root node). This process is repeated at every non-root node in the tree until the root node is reached.

\hangafter=1
\setlength{\hangindent}{2.3em}
2) Once the data-driven signal flow has reached the root node, say node $j^{''}$, the vector $\tilde{\mathbf{z}}\left(l\right)$ is obtained by
\begin{equation}
\begin{gathered}
\tilde{\mathbf{z}}\left(l\right)=\check{\mathbf{D}}^{-1}_{j^{''}}\left(l\right)\check{\mathbf{z}}_{j^{''}}\left(l\right),\\
\check{\mathbf{D}}_{j^{''}}\left(l\right)=\mathbf{D}\left(l\right),\check{\mathbf{z}}_{j^{''}}\left(l\right)=\mathbf{z}\left(l\right).
\end{gathered}
\end{equation}
The vector $\tilde{\mathbf{z}}\left(l\right)$ is now flooded through the WASNs
(away from the root node) so that it reaches every node, where the nodes simple act as relays to pass $\tilde{\mathbf{z}}\left(l\right)$ further through the tree. Finally, the output $d_j\left(l\right)={\mathbf{g}}^H_j\tilde{\mathbf{z}}\left(l\right)$ is obtained.

Based on the data-driven signal flow above, we see that any leaf node will transmit only a single block of compressed signals $\mathbf{z}_q\left(l\right)$. Any non-leaf node will transmit a maximum of two blocks of signals including $\check{\mathbf{z}}_j\left(l\right)$ and $\tilde{\mathbf{z}}\left(l\right)$, first toward the root node and then away from the root node. It is worth noting that $\mathbf{D}_q\left(l\right)$ and $\check{\mathbf{D}}_j\left(l\right)$ only need to be transmitted once for each block of microphone signals and can be ignored.

When the first-order recursive smoothing method is applied, for any leaf node $q$,
$\bar{\mathbf{c}}_q\left(l\right)$ and $c_q\left(l\right)$ are transmitted to reconstruct
$\mathbf{D}_q\left(l\right)$ and $\mathbf{z}_q\left(l\right)$ with (\ref{Reconstruction1}) and (\ref{Reconstruction2}) at the non-root node $j$. The rest is the same as the non-recursive smoothing method. In particularly, $\check{\mathbf{D}}_j\left(l\right)$
needs to be transmitted for each noise-only frame.
\section{Analysis of Complexity and Bandwidth\label{section5}}

This section analyzes the complexity and the bandwith of the proposed beamformer, and those of state-of-the-art beamformers are also presented. For the centralized LCMV/LCMP beamformer, each node needs to have access to the microphone signals from other nodes and therefore all microphone signals in the WASNs are transmitted.
In general, we hope to get the estimation of the $S$ speech sources separately, where the constraint vector $\bar{\mathbf{g}}_j$ should be modified to an $S\times S$ matrix $\bar{\mathbf{G}}_j$ with only one non-zero entry per column.
Without loss of generality, we assume that each node has $M_j=N$ microphones. We need to note that the cost discussed below does not include the overhead associated with those algorithms exploiting a VAD.

The total number of transmissions related to the communication bandwidth is dependent not only on the choice of the beamformer but also on the WASNs topology. As such, it is difficult to analytically bound this transmission cost for any network topology.
The comparison of the communication bandwidth and the computational complexity of different beamformers is performed in a fully connected network, where the centralized LCMV/LCMP, LC-DANSE \cite{LCDANSE}, BD-LCMV/BD-LCMP \cite{DDS2}, and the beamformers proposed in this paper have the same update rate.

We denote the transmission of one real number as one transmission. For the DNBD-LCMV/DNBD-LCMP
with the first-order recursive smoothing method,
each node needs to transmit a $S$-dimensional complex vector $\bar{\mathbf{c}}_j\left(l\right)$ and a real number $c_j\left(l\right)$, where the total number of transmissions is $J\left(2S+1\right)$. Besides, an inversion of an $S\times S$ matrix $\mathbf{D}\left(l\right)$ in (\ref{Multiplier1}) is performed, yielding a complexity of $\mathcal{O}\left(S^3\right)$. For the DNDS,
each node needs to transmit a $S$-dimensional compressed signals $\mathbf{z}_j\left(l\right)$
and the total number of transmissions is $2JS$.
The computational complexity and the communication bandwidth of different beamformers are shown in Table~\ref{Table1}.

\newcommand{\tabincell}[2]{\begin{tabular}{@{}#1@{}}#2\end{tabular}}
\begin{table}[t]
\centering
\scalebox{0.65}{
\begin{threeparttable}
\caption{\large{complexity and bandwidth of different beamformers} \label{Table1}}

\begin{tabular}[b]{c|c|c|c|c|c}
\hline
\multicolumn{2}{c|}{\multirow{2}*{Beamformer}} & \multirow{2}*{Complexity} &\multirow{2}*{Bandwidth} &\multicolumn{2}{c}{\tabincell{c}{$N=6,J=4,S=2$,\\$\lvert \mathcal{L}_y\rvert=100,~t_{\max}=1\tnote{1}$}}\\ \cline{5-6}
\multicolumn{2}{c|}{~}& ~& ~&Complexity & Bandwidth \\
\hline
\multirow{4}{*}{\tabincell{c}{Non-recursive\\ smoothing method}}
&LCMV/LCMP   &$\mathcal{O}\left(\left(JN\right)^3\right)/\lvert \mathcal{L}_y\rvert$  & $2JN$  &$\mathcal{O}\left(13824\right)/100$ & 48\\
\cline{2-6}
 & LC-DANSE  &$\mathcal{O}\left(\left(N+\left(J-1\right)S\right)^3\right)/\lvert \mathcal{L}_y\rvert$ & $2JS$ &$\mathcal{O}\left(1728\right)/100$ & 16\\
\cline{2-6}
& \tabincell{c}{BD-LCMV/\\BD-LCMP} & $\mathcal{O}\left(N^3\right)/\lvert \mathcal{L}_y\rvert$ & $2JS$ & $\mathcal{O}\left(216\right)/100$ & 16\\
\cline{2-6}
& \tabincell{c}{DNBD-LCMV/\\DNBD-LCMP} & $\mathcal{O}\left(S^3\right)/\lvert \mathcal{L}_y\rvert$ & $2JS$ & $\mathcal{O}(8)/100$ & 16\\
\hline
\multicolumn{2}{c|}{DNDS} & $\mathcal{O}(S^3)$ (once)\tnote{2} & $2JS$  & $\mathcal{O}\left(8\right)$(once)\tnote{2} & 16\\
\hline
\multirow{3}{*}{\tabincell{c}{First-order recursive\\ smoothing method}}
 & LCMV/LCMP   &  $\mathcal{O}\left(\left(JN\right)^3\right)$ & $2JN$ & $\mathcal{O}\left(13824\right)$ & 48\\
\cline{2-6}
 & \tabincell{c}{BD-LCMV/\\BD-LCMP} & $\mathcal{O}\left(N^3\right)$ & $2JSt_{\max}\left(S+1\right)$ & $\mathcal{O}\left(216\right)$  & 48\\
\cline{2-6}
& \tabincell{c}{DNBD-LCMV/\\DNBD-LCMP} & $\mathcal{O}\left(S^3\right)$  & $J\left(2S+1\right)$ & $\mathcal{O}\left(8\right)$ & 20\\
\hline
\end{tabular}
  \begin{tablenotes}
  \item[1] $t_{\max}$ is the maximum number of iterations for the BD-LCMV/BD-LCMP.
  \item[2] the weight vector $\bar{\mathbf{w}}^{''}_j$ in (\ref{DNDS}) only needs to be calculated once at the beginning.
  \end{tablenotes}
\end{threeparttable}}
\end{table}

From Table~\ref{Table1}, first, for the beamformers with the first-order recursive smoothing method, the DNBD-LCMV/DNBD-LCMP has the lowest complexity and bandwidth. For the beamformers with the non-recursive smoothing method, the DNBD-LCMV/DNBD-LCMP has the lowest complexity, and has the same bandwidth as the LC-DANSE and the BD-LCMV/BD-LCMP. Second, the bandwidth of the DNDS is not higher than other beamformers. In particular, its complexity is negligible because $\bar{\mathbf{w}}^{''}_j$ in (\ref{DNDS}) only needs to be calculated once at the beginning.
Finally, the complexity of the proposed beamformers and the BD-LCMV/BD-LCMP is independent of the number of nodes $J$ and they are completely scalable, where
there is no increase in the per-node complexity when new nodes are added to the networks.

\section{Experimental Results\label{section6}}
This section evaluates the performance of the proposed beamformers including DNBD-LCMV/DNBD-LCMP and DNDS and compares them with
three state-of-the-art beamformers including centralized LCMV/LCMP and LC-DANSE \cite{LCDANSE}) by using three objective measures, which are SNR, short-time objective intelligibility (STOI) \cite{STOI}, and average TDOA error (ATE) of the speaker in a simulated room when the column space $\mathbf{Q}$ and VAD have errors for reverberation times $T_{60}=0.3~\mathrm{s}$ and $T_{60}=0.5~\mathrm{s}$. The room impulse responses (RIRs) are generated by the image method \cite{ImageMethod,URL1}.

\subsection{Experimental Setup\label{Section61}}
The dimensions of the simulated room is $5~\mathrm{m} \times 5~\mathrm{m} \times 3~\mathrm{m}$ with
reverberation times $T_{60}=0.3~\mathrm{s}$ and $T_{60}=0.5~\mathrm{s}$.
The WASNs consist of $J=4$ nodes, each having $M_j=6$ microphones forming a uniform linear array with an inter-microphone distance of 3 cm. Four point sound sources including $S=2$ speech sources and two babble noise sources are presented. The configuration of the nodes and sound sources are depicted in Fig.~\ref{ExperimentSetup}.
The two speakers that produce speech sentences taken from the NOIZEUS corpus \cite{Corpus} have the same power. The two babble directional noise sources are mutually uncorrelated with power that is $10\%$ of the power of any speaker, i.e., 10 dB SNR. Besides the two babble directional noise sources, all microphone signals have an uncorrelated white Gaussian noise component with 30 dB SNR with respect to the superimposed speech signals in the first microphone of node 1. The sampling frequency is 8 kHz and the sound speed is $c=343~\mathrm{m/s}$.
\begin{figure}[t]
\centering
\includegraphics[width=70mm]{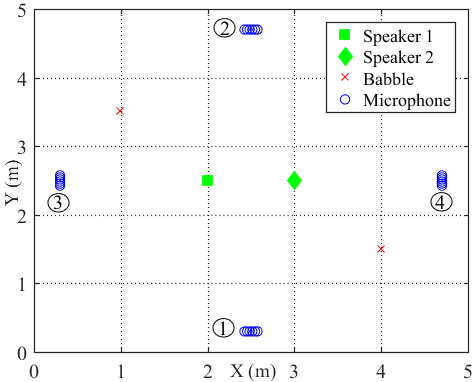}
\caption{The acoustic scenario used in the experiment. There are $S=2$ speakers, two babble directional noise sources, and $J=4$ nodes with $M_j=6$ microphones each.
 The nodes are located at the center of each of the four walls, 30 cm from the walls. All nodes and all sources are in the same horizontal plane, 1.5 m above ground level.}\label{ExperimentSetup}
\end{figure}

In order to approximate the real acoustic scenario, these positions of the speakers, where the column space $\mathbf{Q}$ was estimated, were uniformly distributed over a sphere centered around the true source positions depicted in Fig.~\ref{ExperimentSetup}.
These positions were referred to as training positions. For Speaker 1 and Speaker 2, its erroneous training positions had radii $r_1$ and $r_2$ ranging from 0 to 10 cm, respectively.
Therefore, the column space estimation error can be modeled as a function of positional error between the training positions and the true source positions \cite{DDS2}. For every value of the positional error, the average performance of 100 different setups were measured. Each setup used the same source signals at the true source positions. However, a different set of training positions mentioned previously and different realizations of the microphone-self noise were used in each setup.

In the experiment, the non-recursive smoothing method was adopted to estimate $\mathbf{\Delta}_{yy,j}$ and $\mathbf{\Delta}_{nn,j}$, and the estimation was performed on the
entire length of signal \cite{TIDANSE,DetectionError},
\begin{equation}
\begin{gathered}
\mathbf{\Delta}_{yy,j}=\frac{1}{\lvert \mathcal{L}_y \rvert}\sum_{l\in \mathcal{L}_y}\mathbf{y}_j(l)\mathbf{y}^H_j(l),\\
\mathbf{\Delta}_{nn,j}=\frac{1}{\lvert \mathcal{L}_y \rvert}(\sum_{l_1\in \mathcal{L}_y\setminus \mathcal{L}^{'}_y}\mathbf{n}_j(l_1)\mathbf{n}^H_j(l_1)+\sum_{l_2\in \mathcal{L}^{'}_y}\mathbf{y}_j(l_2)\mathbf{y}^H_j(l_2)),
\end{gathered}
\end{equation}
where $\mathcal{L}_y$ is the set of frames of the entire time horizon, and $\mathcal{L}^{'}_y$ is the set of frames of the noisy signals used to estimate $\mathbf{\Delta}_{nn,j}$. The set $\mathcal{L}^{'}_y$ is used to simulate the VAD error, where the noisy frames containing speech component are erroneously detected as noise-only frames. The error can be measured by the following scalar
\begin{equation}
R=\frac{\lvert \mathcal{L}^{'}_y \rvert}{\lvert \mathcal{L}_y\rvert}\times 100\%.
\end{equation}
When $\mathcal{L}^{'}_y$ is an empty set, i.e., $\mathcal{L}^{'}_y=\{\varnothing\}$ and $R=0$, an ideal VAD is considered.

The SNR is the ratio between the powers of the desired speech component and the noise, and can be defined as
\begin{equation}
\mathrm{SNR}=10\log{\frac{E\{\bar{d}^2_j(t)\}-E\{\bar{n}^2_j(t)\}}{E\{\bar{n}^2_j(t)\}}},
\end{equation}
where $\bar{d}_j(t)$ and $\bar{n}_j(t)$ denote the time domain beamformer output and the time domain noise component (also containing the residual competing speech component \cite{DGSC}) at the $j$th node, respectively.

The TDOA error $\Delta {T}_{j1,s}$ between the $j$th node and the 1st node for the $s$th speaker can expressed by
\begin{equation}
\begin{gathered}
\Delta {T}_{j1,s}=\begin{cases}
T_{j1,s}-\hat{T}_{j1,s}, & \mathrm{if} ~T_{j1,s}-\hat{T}_{j1,s} \geq 0;\\
\hat{T}_{j1,s}-T_{j1,s}, & \mathrm{if} ~T_{j1,s}-\hat{T}_{j1,s} < 0.
\end{cases}\\
T_{j1,s}=\left(\lVert \mathbf{x}_{j1}-\bar{\mathbf{x}}_{s}\rVert-\lVert \mathbf{x}_{11}-\bar{\mathbf{x}}_{s}\rVert\right)/c,
\end{gathered}
\end{equation}
where $\mathbf{x}_{j1}$ is the location of the 1st microphone at the $j$th node, $\bar{\mathbf{x}}_s$ is the location of the $s$th speaker, and $\lVert \cdot \rVert$ denotes the two-norm of a vector. $T_{j1,s}$ is the theoretical TDOA and $\hat{T}_{j1,s}$ is the estimated TDOA based on the output signals of different nodes using the
generalized cross-correlation phase transform (GCC-PHAT) \cite{GCC}. More details about TDOA can be found in \cite{TDOA}. The ATE is defined by
\begin{equation}
\mathrm{ATE}=\frac{1}{J-1}\sum_{j=2}^J \Delta {T}_{j1,s}.
\end{equation}

For the following performance comparison, the first microphone of each node is chosen as the reference microphone and the experimental results for Speaker 1 are presented. In particularly, LC-DANSE \cite{LCDANSE} is time-recursive and converges to the solution of its centralized algorithm. Therefore, LC-DANSE is replaced by the centralized LCMV \cite{DDS2}.

\subsection{Robustness to Column Space Estimation Error\label{Section62}}
Fig.~\ref{Simulation1} shows the performance of different beamformers under different positional errors in terms of reverberation time $T_{60}=0.3$ s and VAD error $R=5\%$. The spectrograms of the desired signal received by the first microphone of node 1 and the output signals of node 1 of different beamformers were depicted in Fig.~\ref{Simulation2} for $T_{60}=0.3$ s, $R=5\%$, and $r_1=r_2=5$ cm.
\begin{figure}[t]
\centering
\subfigure[]{
\includegraphics[width=38mm]{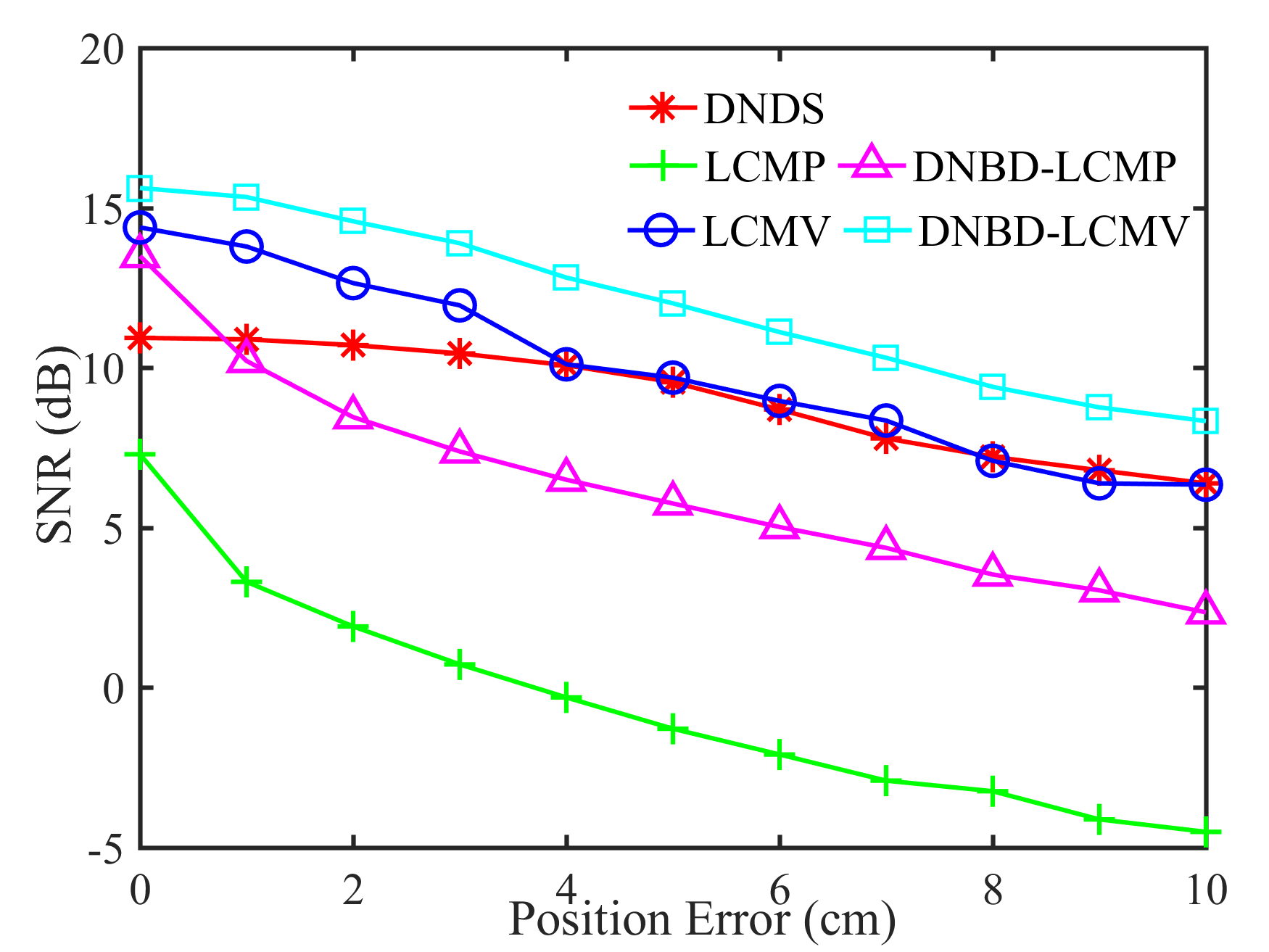}}
\subfigure[]{
\includegraphics[width=38mm]{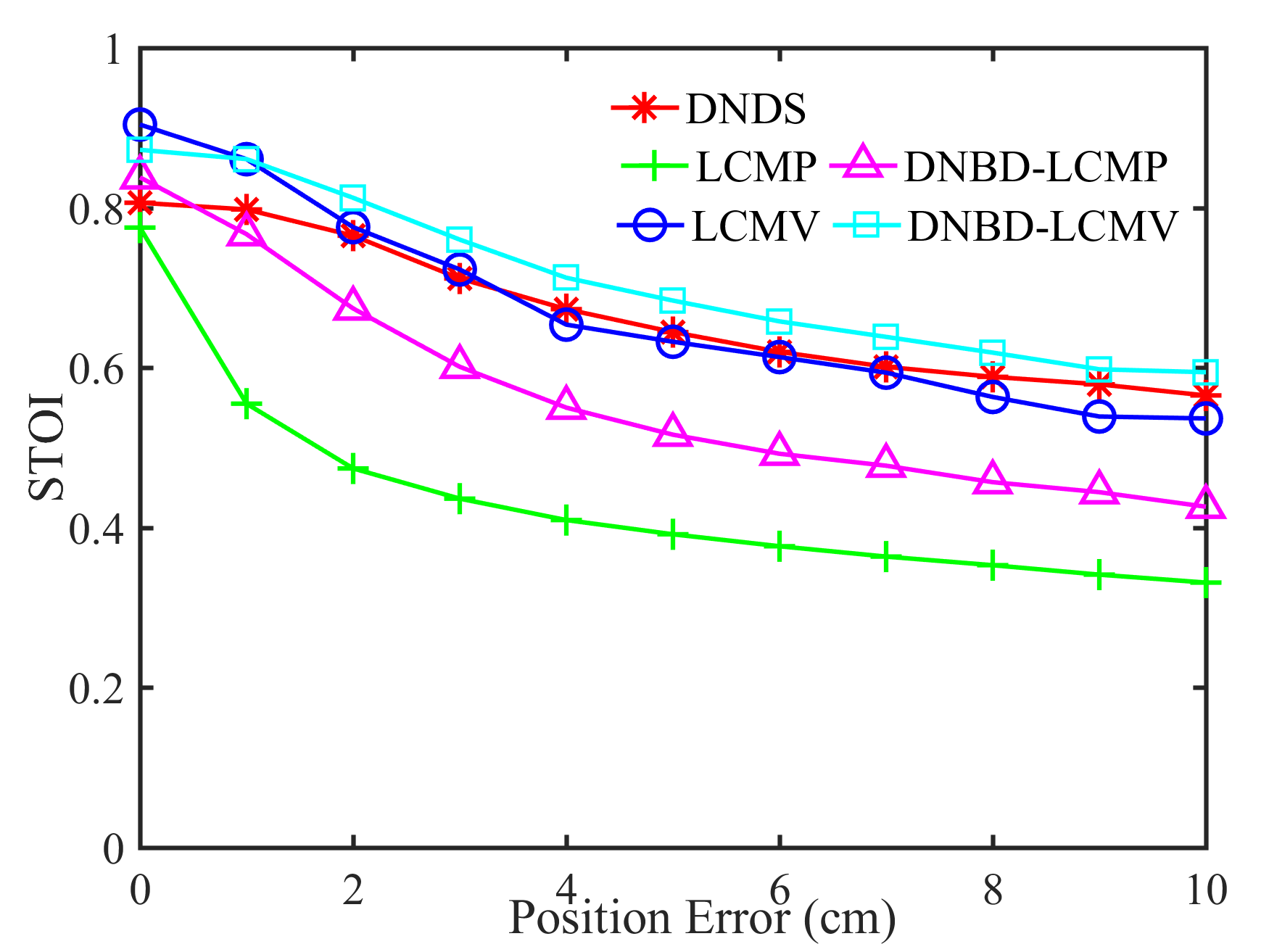}}
\subfigure[]{
\includegraphics[width=38mm]{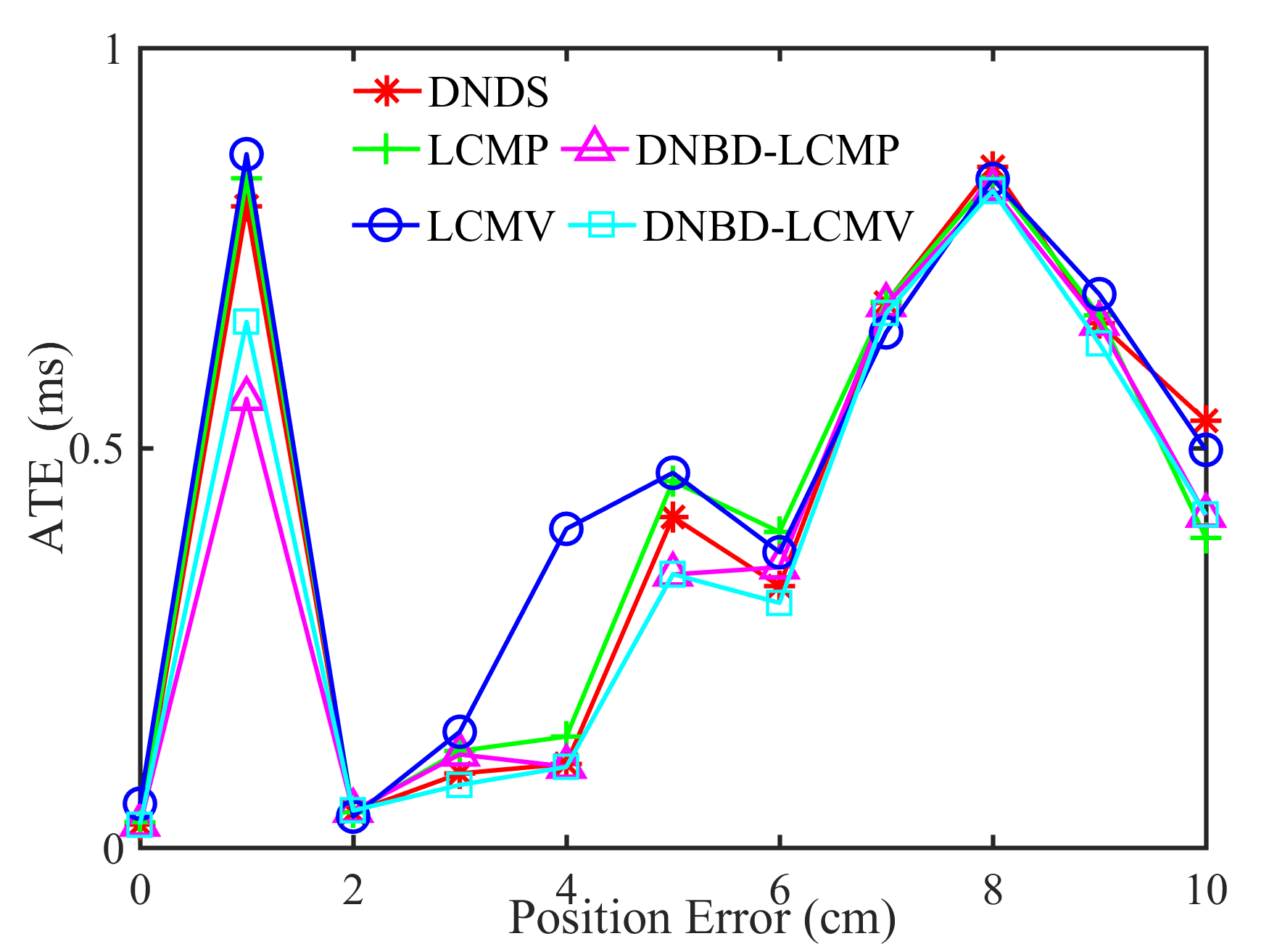}}
\caption{Comparison of SNR, STOI, and ATE for different beamformers ($T_{60}=0.3$ s and $R=5\%$).  (a)-(b) SNR and STOI of output signal of node 1; (c) ATE of output signals of different nodes.\label{Simulation1}}
\end{figure}

From Fig.~\ref{Simulation1}, first, one can get that the performance of the five beamformers decreases
with increasing positional error. When there is no positional error, i.e., $r_1=r_2=0$ cm, the SNR and the STOI of DNBD-LCMP are higher than LCMP and DNDS, and are slightly lower than LCMV and DNBD-LCMV. When there is positional error,
the SNR and the STOI of DNBD-LCMP are significantly reduced, and are much lower than DNDS, LCMV, and DNBD-LCMV.
For different positional error values, the SNR and the STOI of LCMP are the lowest. LCMP and DNBD-LCMP had lower robustness to the positional error than DNDS, LCMV, and DNBD-LCMV. Second, for DNDS, LCMV, and DNBD-LCMV, the SNR and the STOI of DNDS are the lowest when the positional error is zero and then approach LCMV and DNBD-LCMV with increasing positional error (even higher than LCMV).
This is related to the fact that DNDS cannot well suppress the directional noise sources that are not included in the desired response vector but has more robust performance to the positional error \cite{robust}. DNBD-LCMV has higher SNR and STOI than LCMV for non-zero positional error and is less sensitive to the positional error, where $\mathbf{\Delta}_{nn,j}$ has lower dimensions than the full-element noise sample covariance matrix used in the LCMV and is numerically more favorable. Third, the ATEs of different beamformers do not appear to be a significant difference.

\begin{figure*}[t]
\centering
\subfigure[]{
\includegraphics[width=38mm]{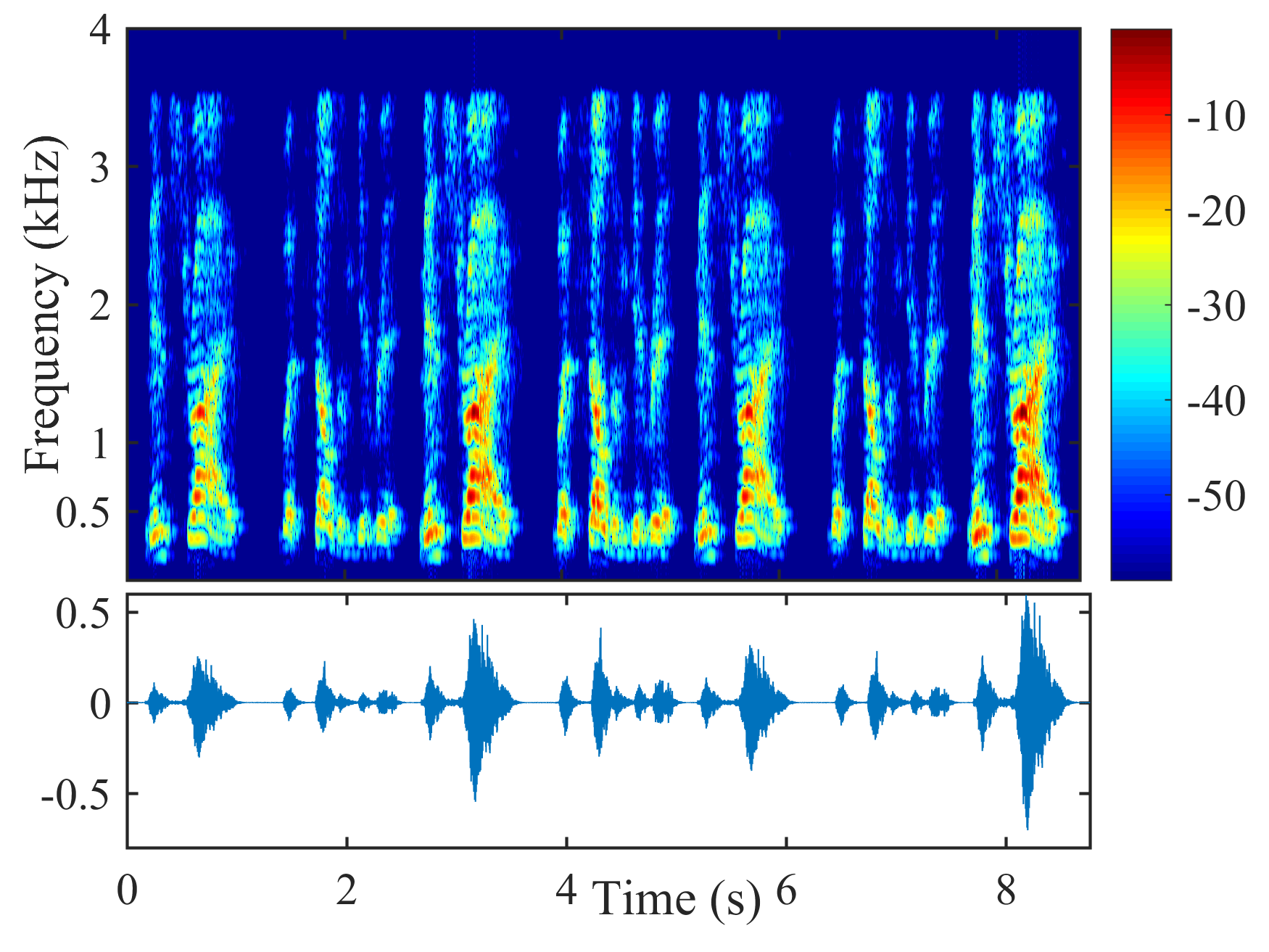}}
\subfigure[]{
\includegraphics[width=38mm]{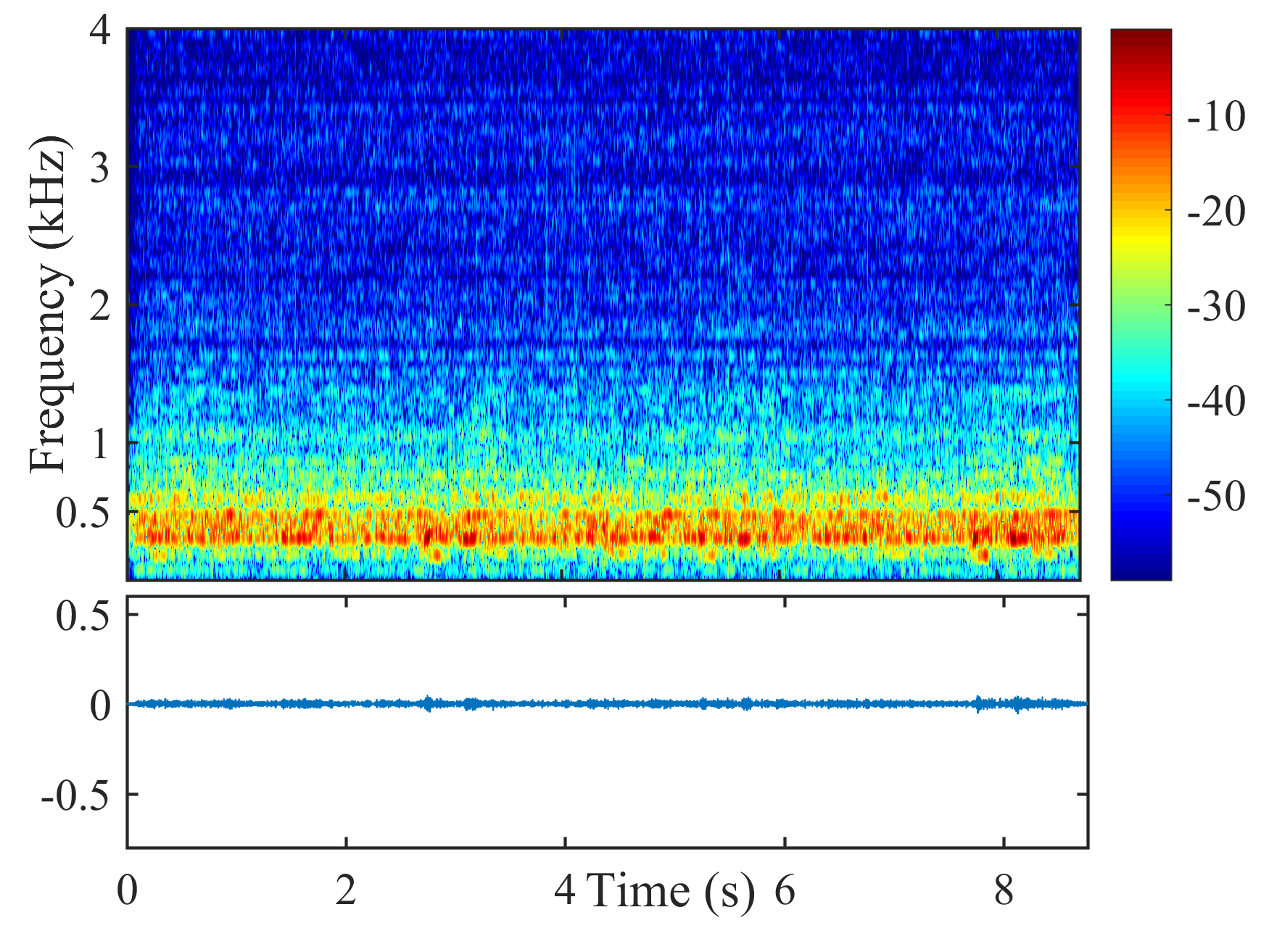}}
\subfigure[]{
\includegraphics[width=38mm]{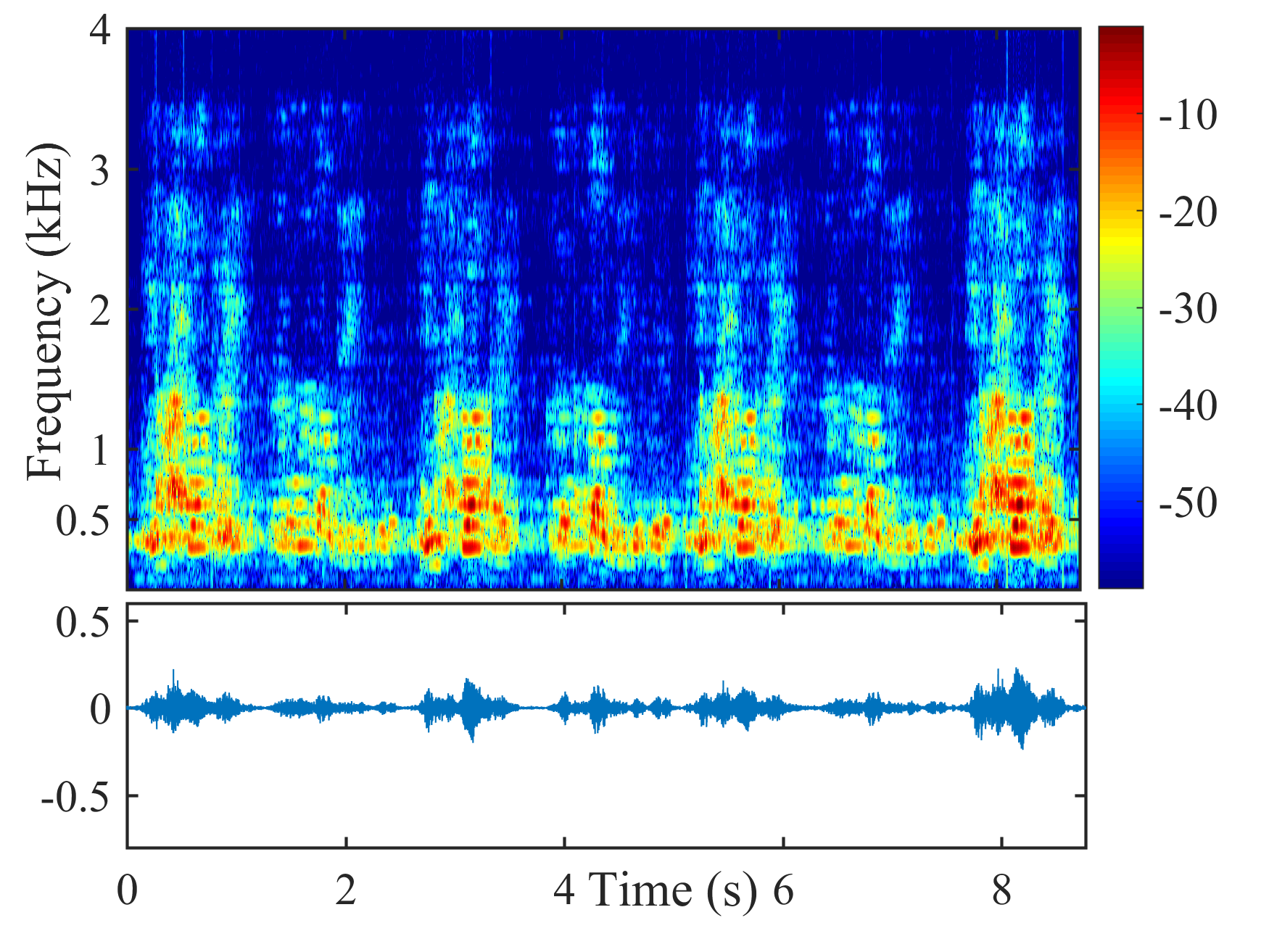}}
\subfigure[]{
\includegraphics[width=38mm]{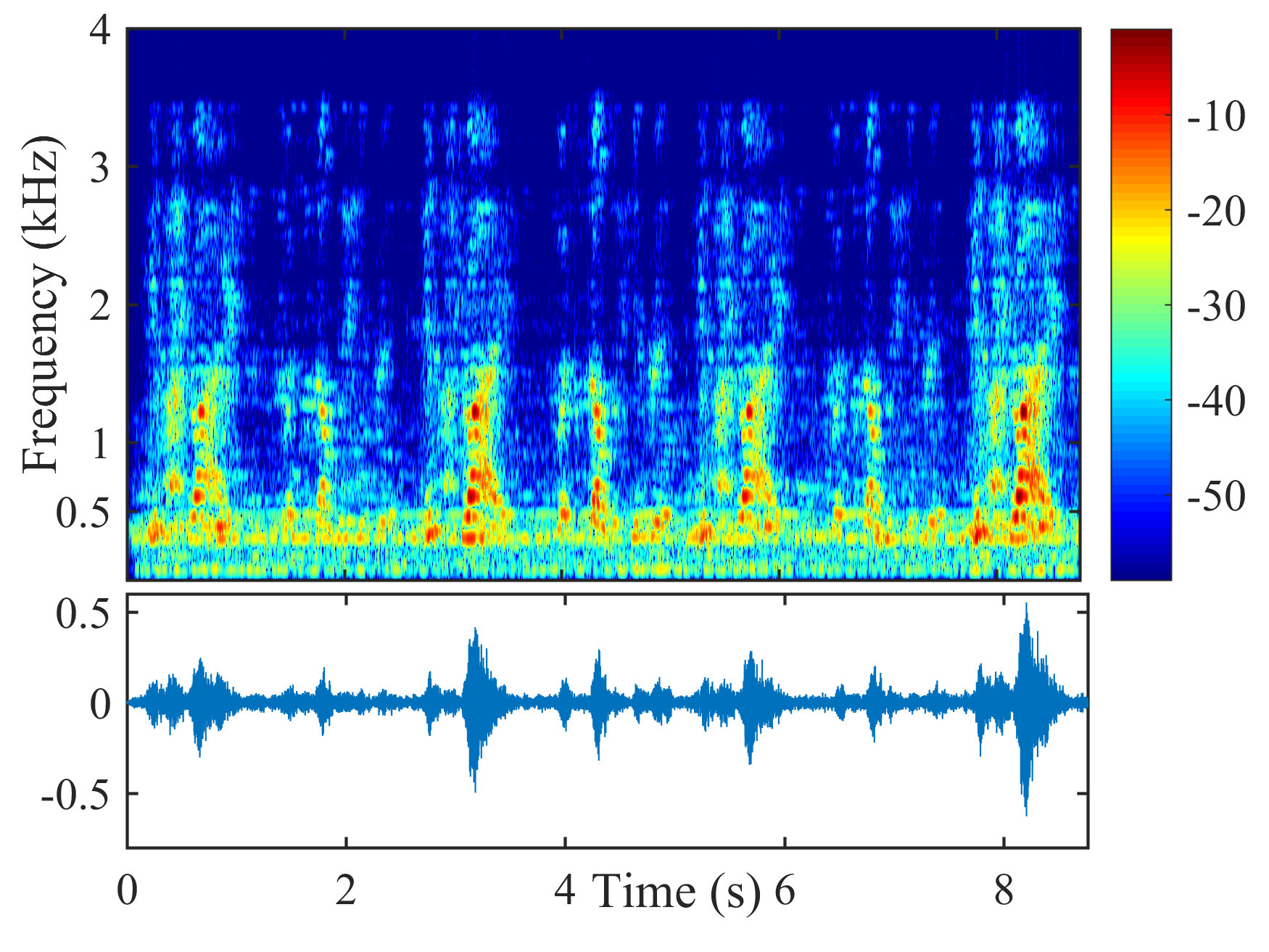}}
\subfigure[]{
\includegraphics[width=38mm]{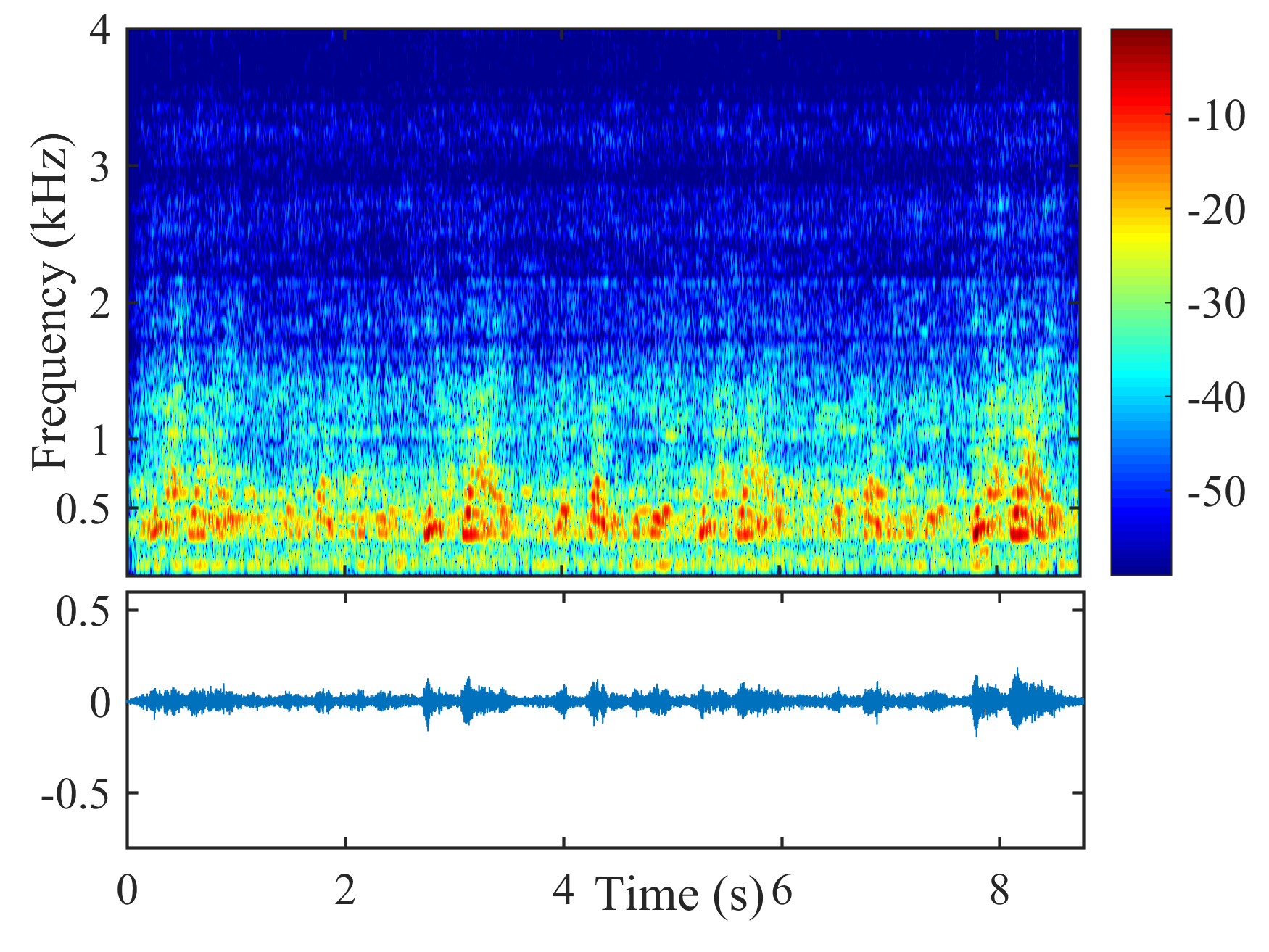}}
\subfigure[]{
\includegraphics[width=38mm]{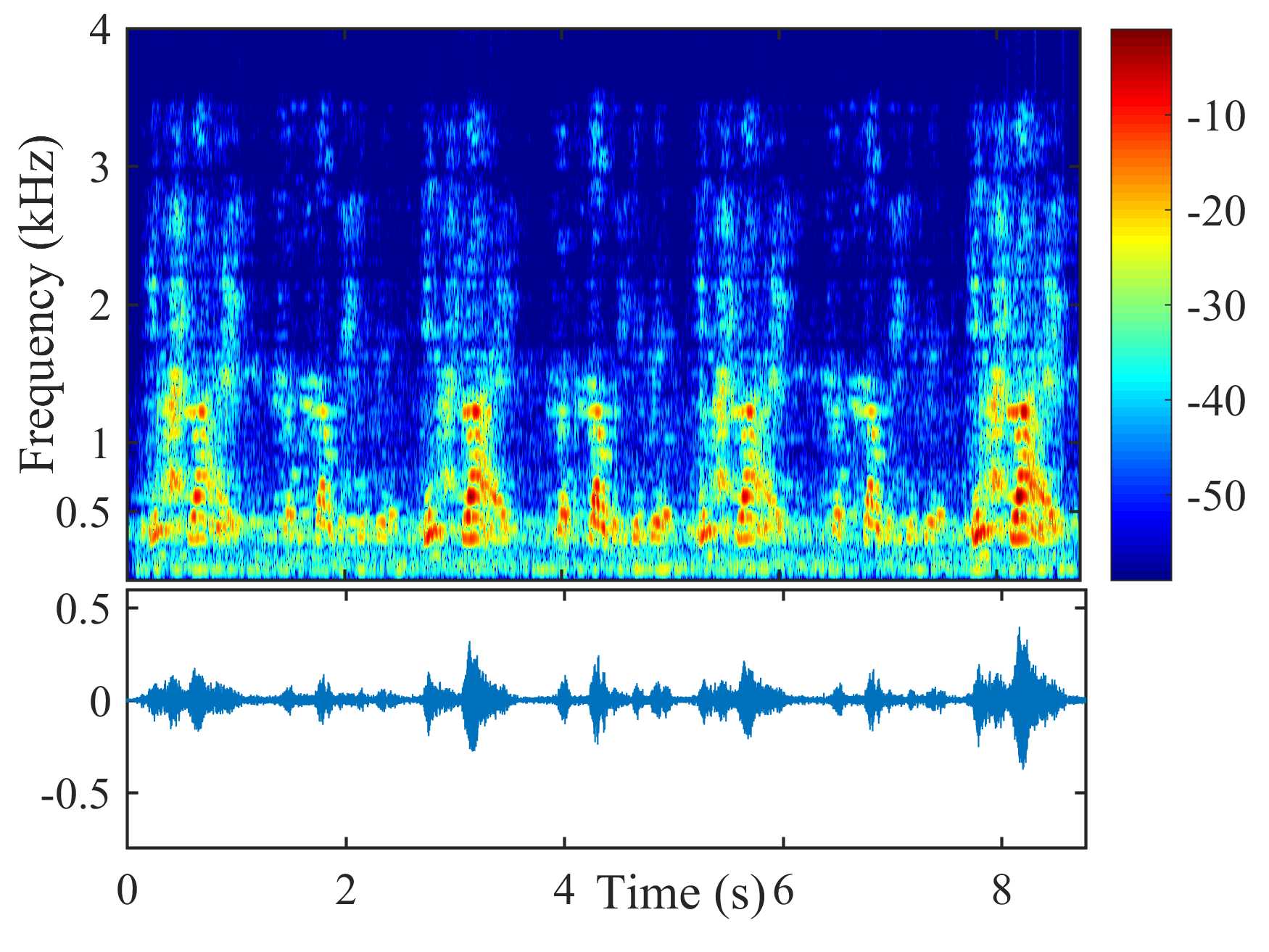}}
\caption{The spectrograms of the desired signal received by the 1st microphone of node 1, and the output signals of node 1 of different beamformers for $T_{60}=0.3$ s, $R=5\%$, and $r_1=r_2=5$ cm. (a) Desired signal; (b) LCMP; (c) LCMV; (d) DNDS; (e) DNBD-LCMP; (f) DNBD-LCMV.\label{Simulation2}}
\end{figure*}

From Fig.~\ref{Simulation2}, the speech component in the output signal of LCMP is almost completely removed, where the column space estimation error results in removal of the actual speech component and preservation in the direction of the wrongly estimated column space.
 NBD-LCMP has the similar problem to LCMP. However, the performance degradation was not that great as with LCMP, this is because DNBD-LCMP has lower degrees of freedom to satisfy the wrong distortionless response and suppresses less speech component than LCMP \cite{DDS2}. Note that LCMP and DNBD-LCMP will not be further considered in the following experiments due to their poor performance in signal model mismatch cases.
LCMV causes more speech distortion than DNBD-LCMV. DNDS preserved the speech component like DNBD-LCMV. However, it also preserves more noise component around 0.5 kHz.

\subsection{Robustness to VAD Error\label{Section63}}

\begin{figure*}[t]
\centering
\subfigure[]{
\includegraphics[width=38mm]{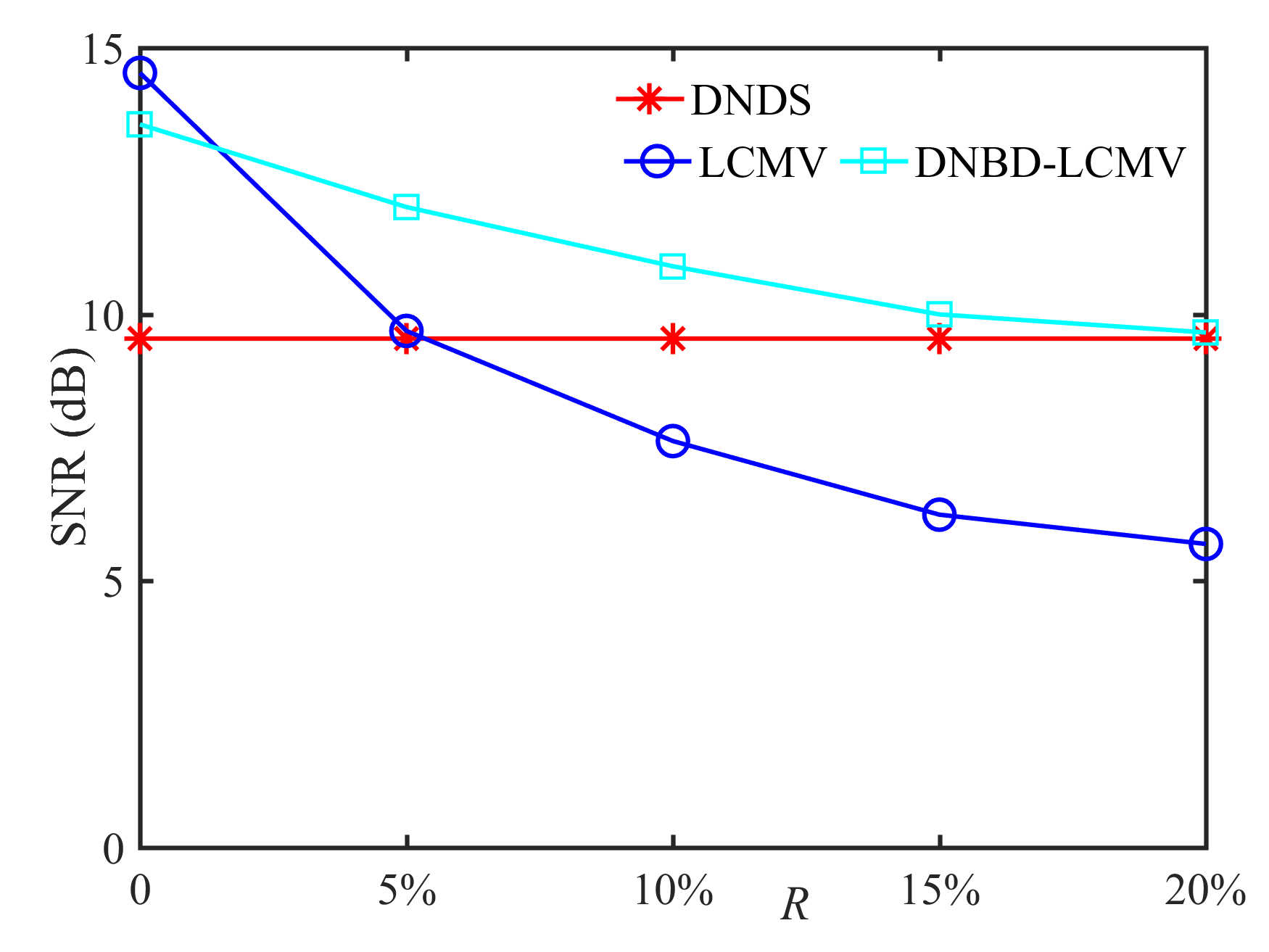}}
\subfigure[]{
\includegraphics[width=38mm]{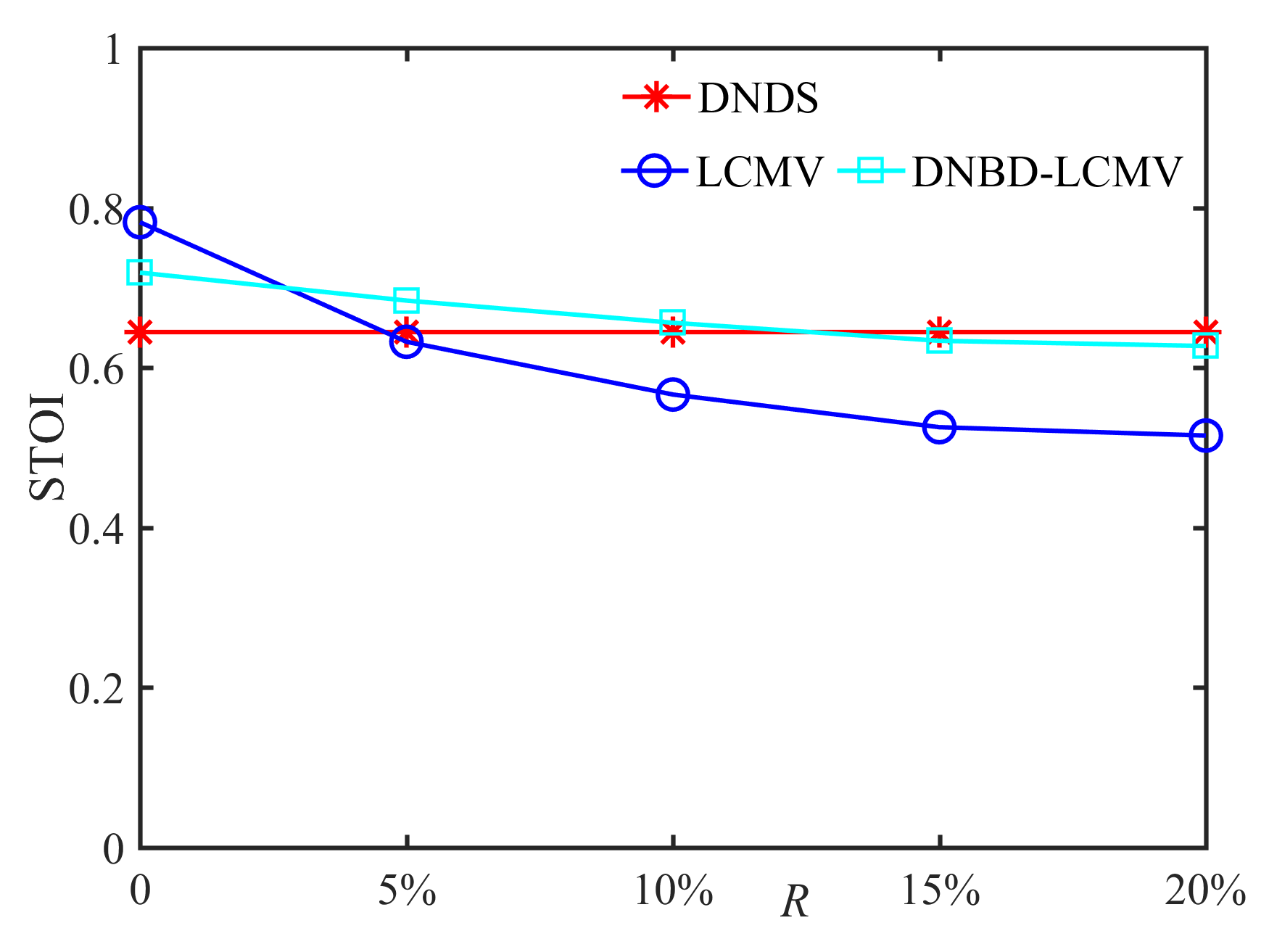}}
\subfigure[]{
\includegraphics[width=38mm]{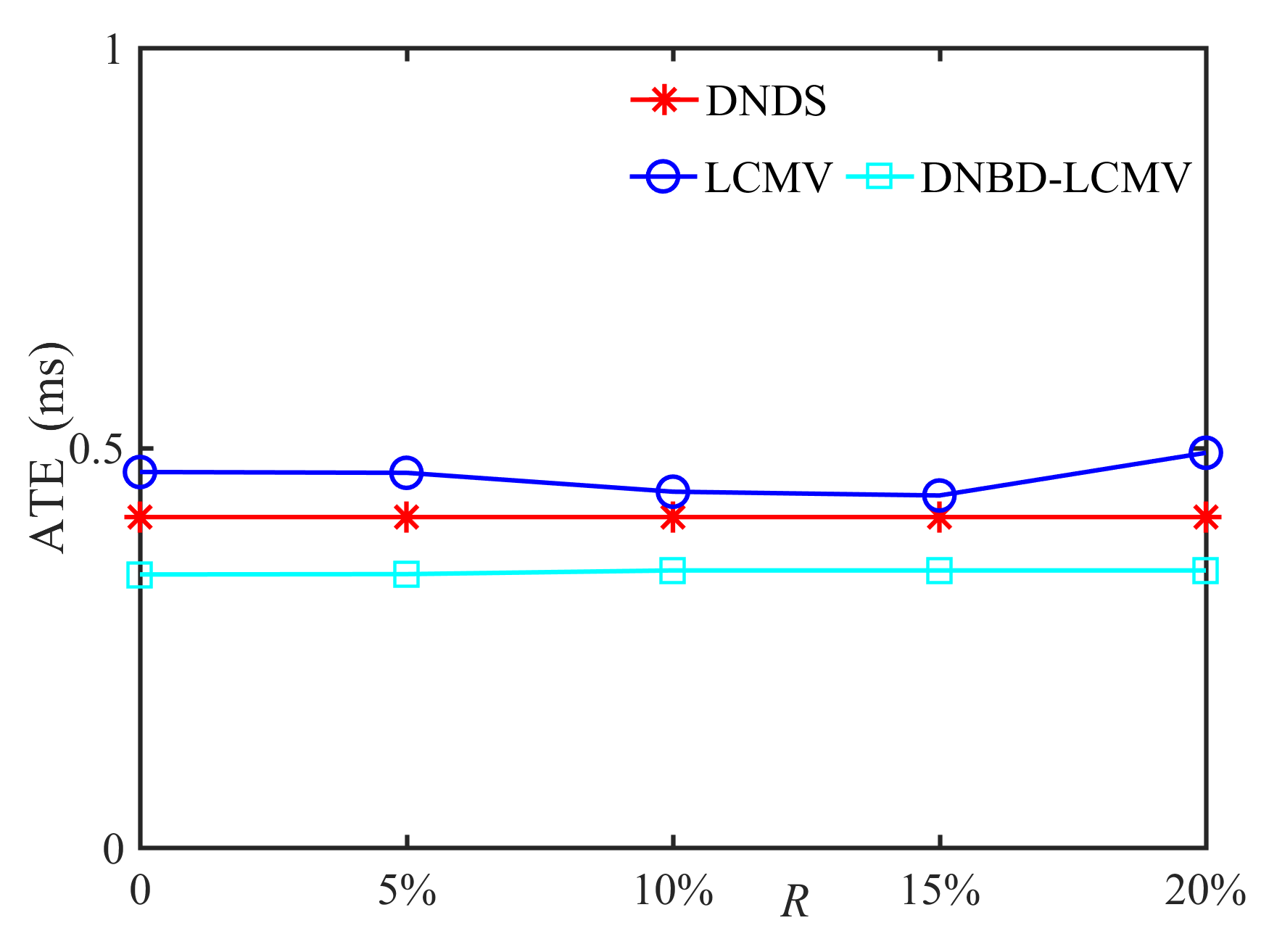}}
\caption{Comparison of SNR, STOI, and ATE for different beamformers ($T_{60}=0.3$ s and $r_1=r_2=5$ cm).  (a)-(b) SNR and STOI of output signal of node 1; (c) ATE of output signals of different nodes.\label{VAD}}
\end{figure*}

In this subsection, we compare the performance of DNDS, LCMV, and DNBD-LCMV under different VAD errors with the reverberation time $T_{60}=0.3$ s and the positional error $r_1=r_2=5$ cm, as shown in Fig.~\ref{VAD}. First, for an ideal VAD, i.e., $R=0$, the LCMV has the highest SNR and STOI.
However, VAD error is often inevitable for practical applications.
The SNR and the STOI of LCMV deteriorate significantly and become the lowest with increasing VAD error, which indicates that LCMV is the most sensitive to the VAD error.
Second, when the VAD error becomes larger, the SNR and the STOI of DNBD-LCMV are reduced and gradually approach the DNDS, which is independent of the VAD error. Third, for different beamformers, their ATEs are almost identical and do not become larger with increasing VAD error.

\subsection{Robustness to Reverberation Time\label{Section64}}
In this subsection, we compare the performance of DNDS, LCMV, and DNBD-LCMV
when the reverberation time increases to $T_{60}=0.5$ s, as shown in Fig.~\ref{Reverberant}. First, one can observe that the SNR and the STOI of LCMV are the lowest for non-zero positional error and the VAD error.
The SNR and the STOI of DNBD-LCMV are reduced, and gradually approach or even lower than DNDS with increasing positional error or the VAD error. Second, no significant difference can be observed in the ATEs of different beamformers, which is similar to Fig.~\ref{Simulation1} (c) and Fig.~\ref{VAD} (c).

\begin{figure*}[t]
\centering
\subfigure[]{
\includegraphics[width=38mm]{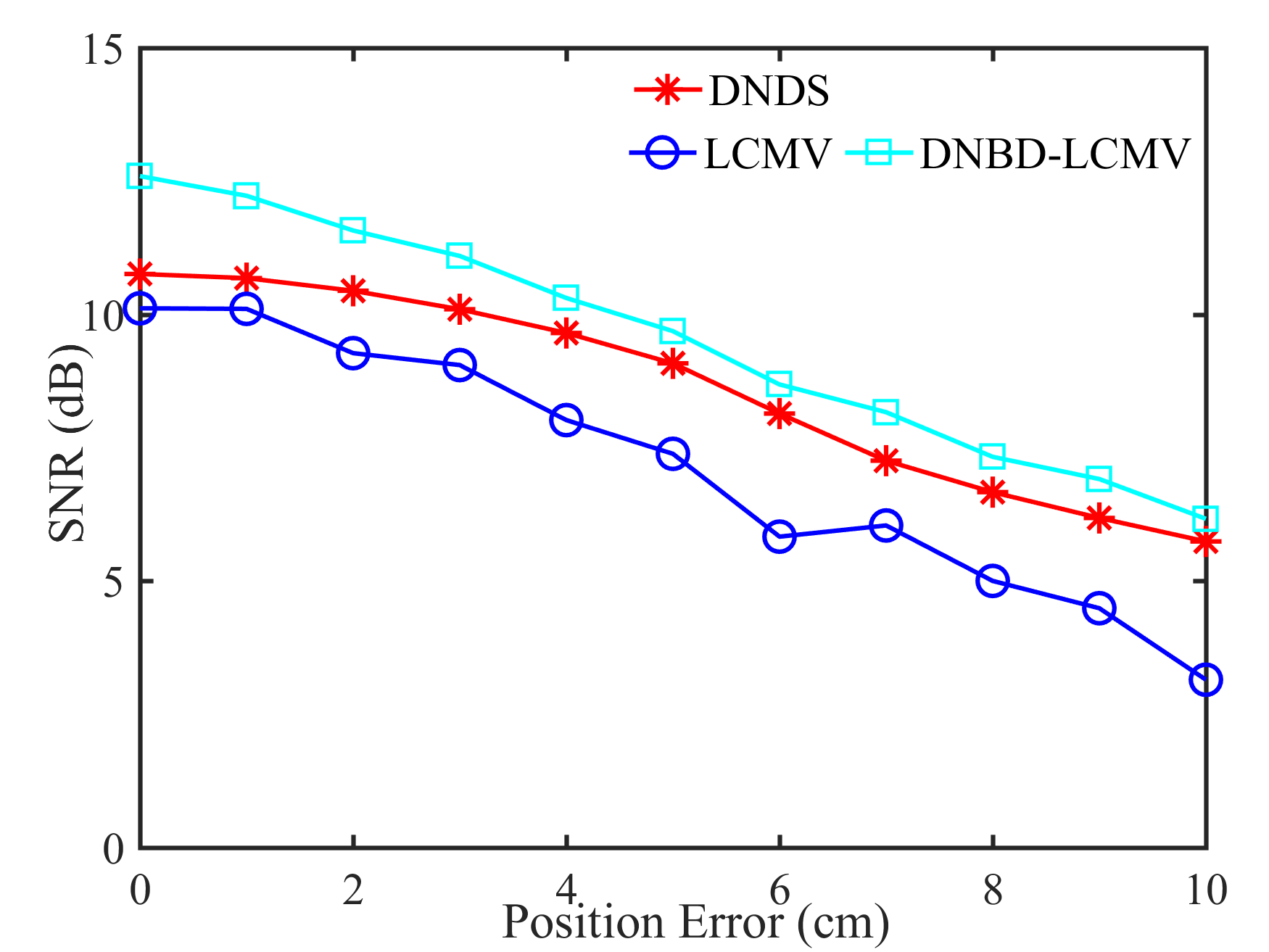}}
\subfigure[]{
\includegraphics[width=38mm]{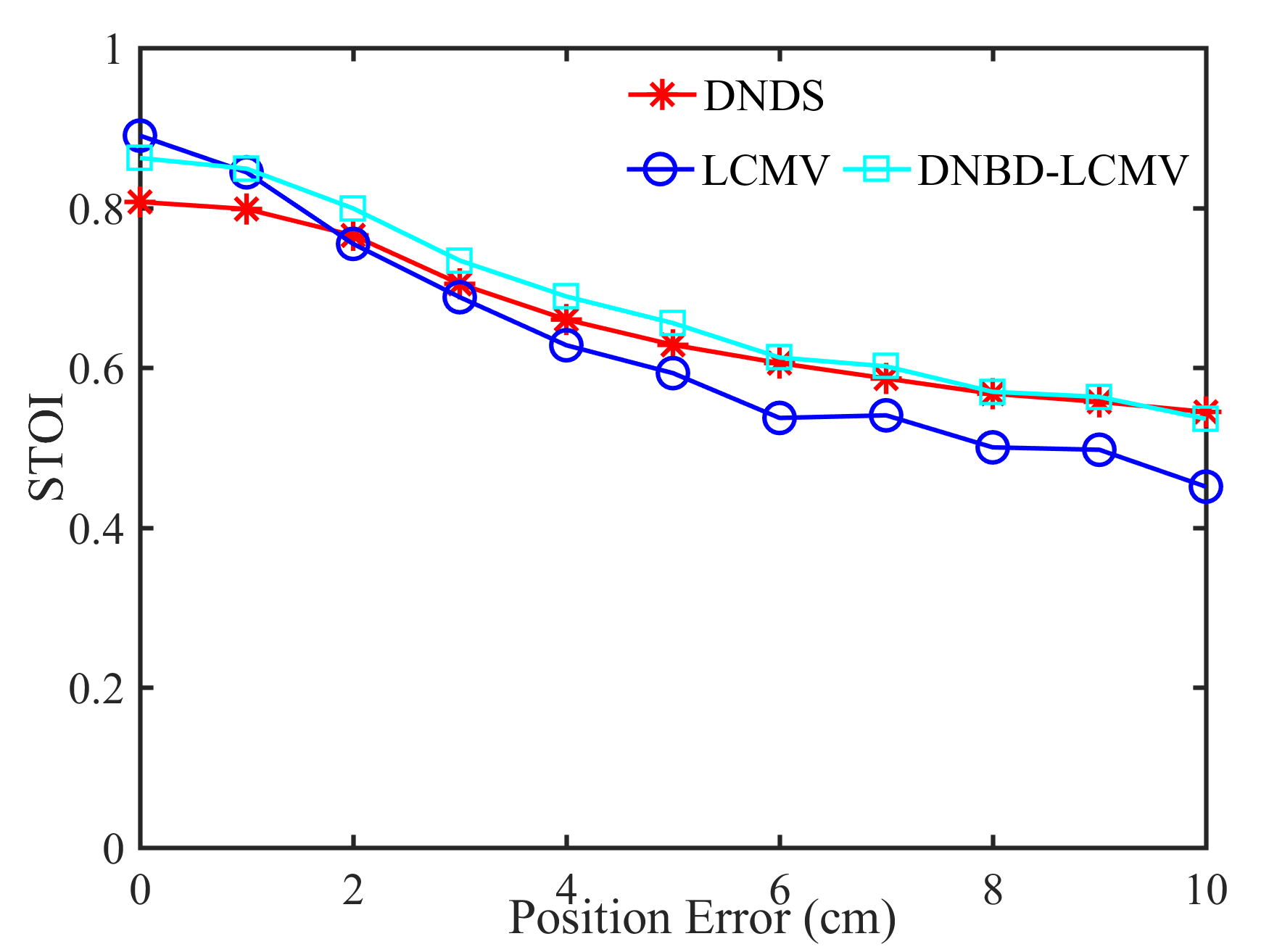}}
\subfigure[]{
\includegraphics[width=38mm]{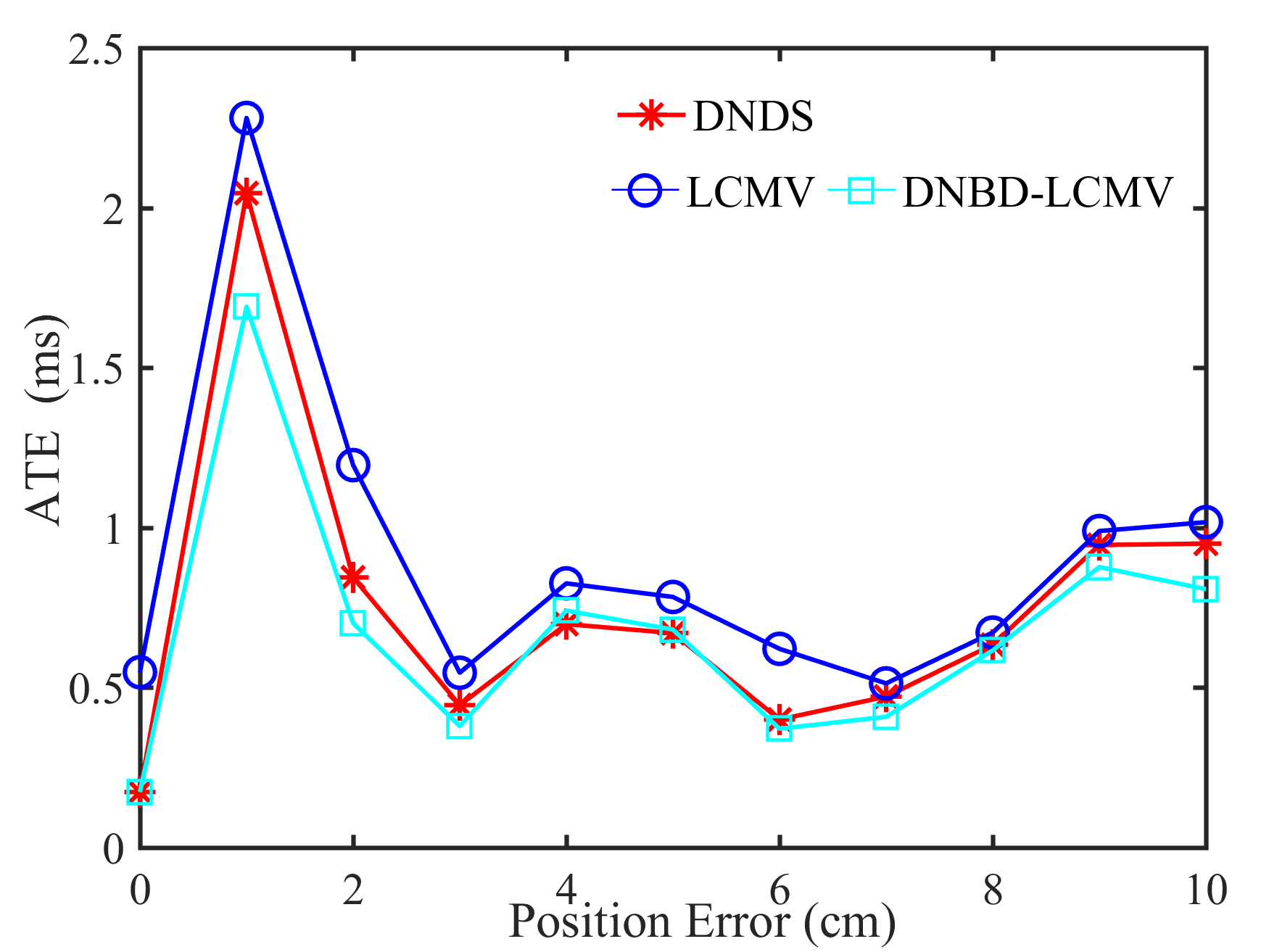}}
\subfigure[]{
\includegraphics[width=38mm]{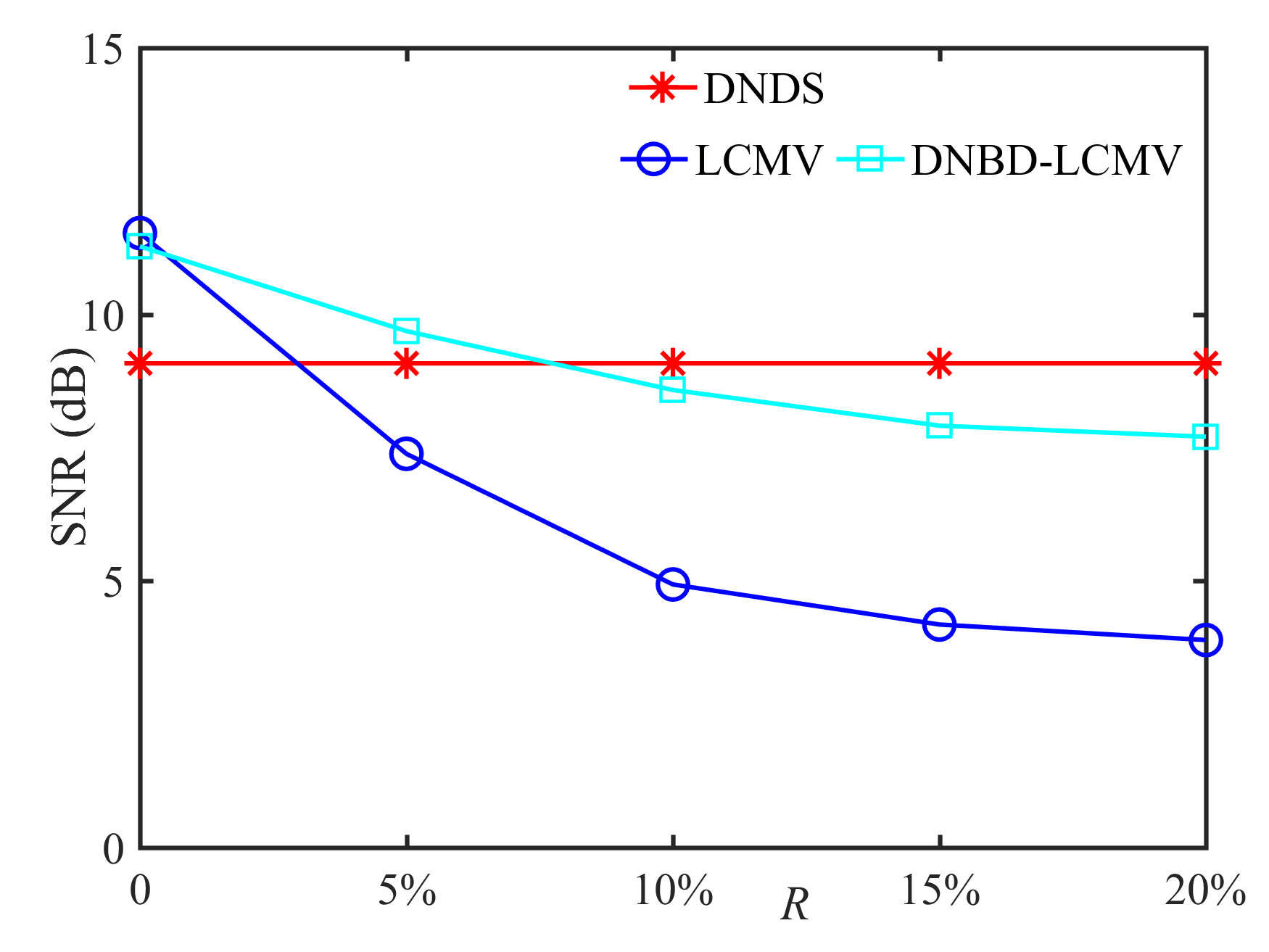}}
\subfigure[]{
\includegraphics[width=38mm]{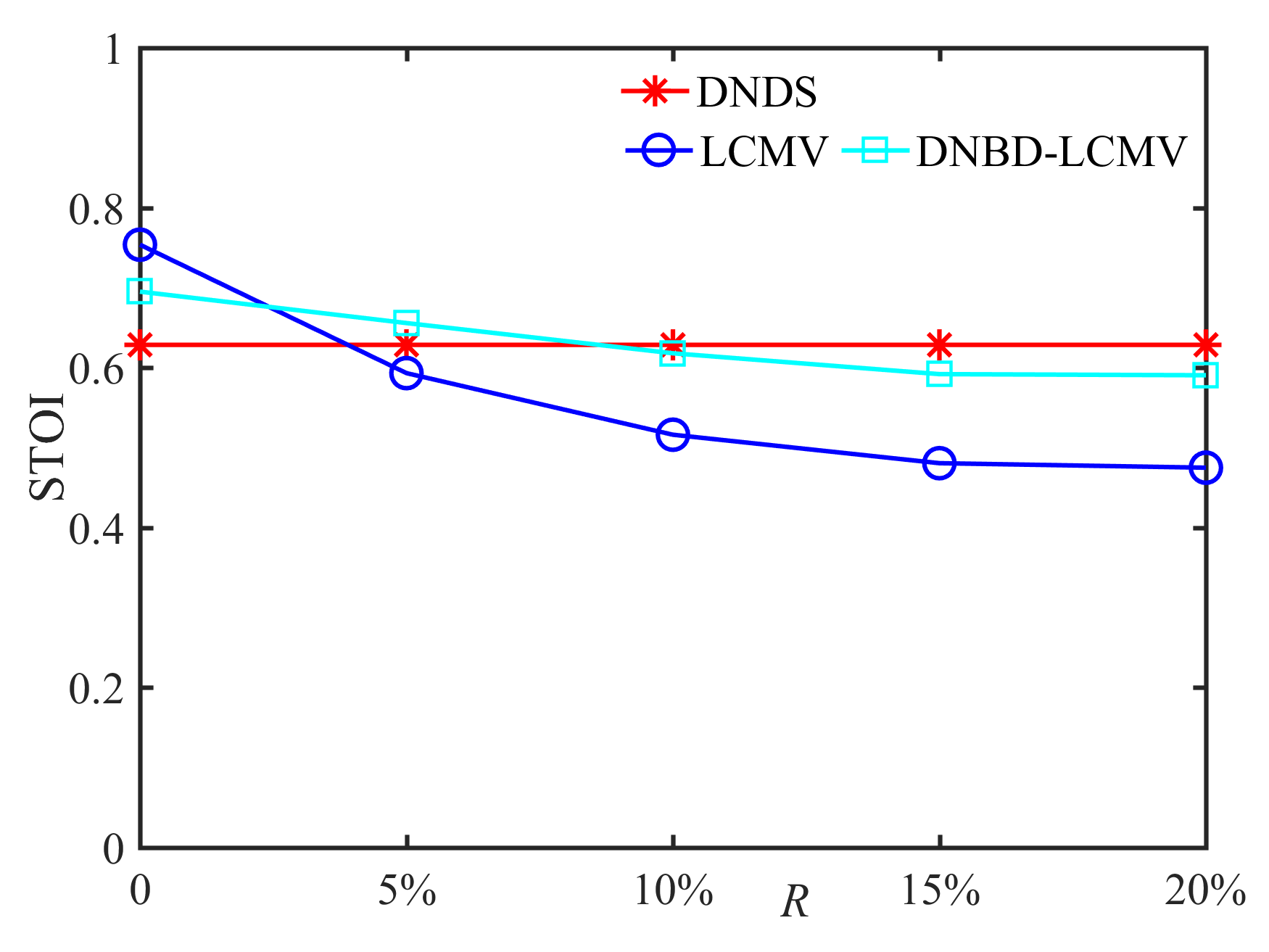}}
\subfigure[]{
\includegraphics[width=38mm]{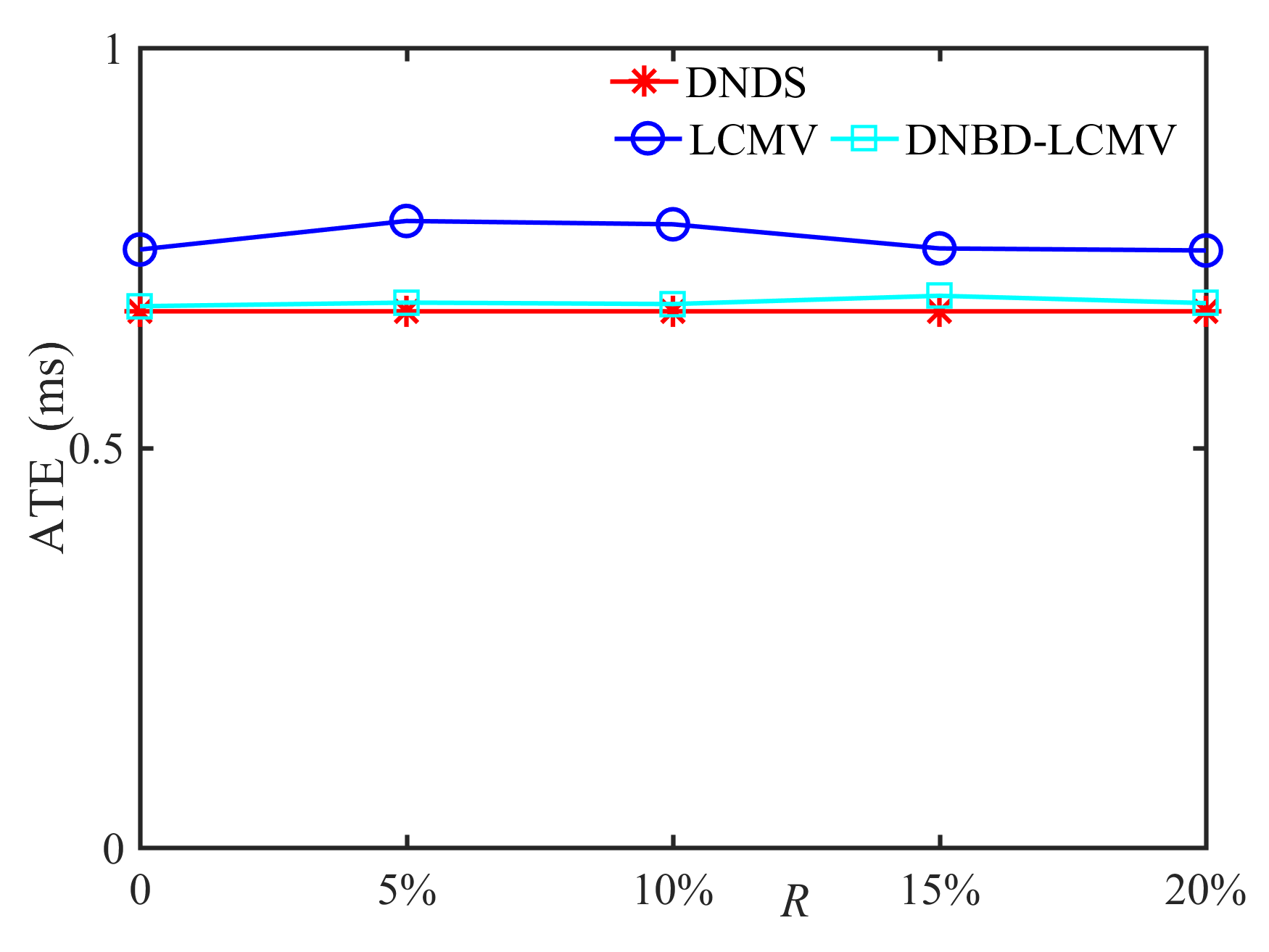}}
\caption{Comparison of SNR, STOI, and ATE for different beamformers ($T_{60}=0.5$ s).  (a)-(c) $R=5\%$; (d)-(f) $r_1=r_2=5$ cm.\label{Reverberant}}
\end{figure*}

From Figs.~\ref{Simulation1}, \ref{VAD}, and \ref{Reverberant}, the performance
including SNR, STOI, and ATE of DNDS, LCMV, and DNBD-LCMV becomes worse for a larger reverberation time, and DNDS and DNBD-LCMV are less sensitive to the reverberation time than LCMV.

\section{Conclusion \label{section7}}
In this paper, we propose a distributed node-specific block-diagonal linearly constrained minimum variance (DNBD-LCMV) beamformer, where the block-diagonal noise covariance matrix is considered to derive its analytical solution from the centralized LCMV beamformer. By updating the inversion of the noise sample covariance matrix using the Sherman-Morrison-Woodbury formula, the exchanged signals can be computed in a more efficient way. The proposed DNBD-LCMV significantly reduces the number of signals exchanged between nodes and exactly solves the LCMV beamformer optimally in each frame.
Analysis and experimental results confirm that the proposed DNBD-LCMV has much lower computational complexity, and is also more robust to column space estimation error and the VAD error than other state-of-the-art distributed node-specific algorithms.

\newpage
\section*{Reference}
\bibliography{StandFormat}

\end{document}